\documentclass[12pt]{article}
\usepackage[totalwidth=460truept,totalheight=638truept]{geometry}
\usepackage{latexsym,graphicx,amsfonts,amssymb,amsmath,xcolor,cancel,MnSymbol}
\usepackage[hypertexnames=false,hidelinks]{hyperref}

\def\theequation{\arabic{section}.\arabic{equation}}

\renewcommand{\theequation}{\thesection.\arabic{equation}}
\global\arraycolsep=1truept
\renewcommand{\theequation}{\arabic{section}.\arabic{equation}}
\begin{document}

\null

\vskip.0truecm

\begin{center}
{\huge \textbf{Purely Virtual Extension}}

\vskip.5truecm

{\huge \textbf{of Quantum Field Theory}}

\vskip.5truecm

{\huge \textbf{for\ Gauge Invariant Fields:}}

\vskip.52truecm

{\huge \textbf{Yang-Mills Theory}}

\vskip.4truecm

\textsl{Damiano Anselmi}

\vskip.1truecm

{\small \textit{Dipartimento di Fisica \textquotedblleft
E.Fermi\textquotedblright , Universit\`{a} di Pisa, Largo B.Pontecorvo 3,
56127 Pisa, Italy}}

{\small \textit{INFN, Sezione di Pisa, Largo B. Pontecorvo 3, 56127 Pisa,
Italy}}

{\small damiano.anselmi@unipi.it}

\vskip.5truecm

\textbf{Abstract}
\end{center}

We extend quantum field theory by including purely virtual\
\textquotedblleft cloud\textquotedblright\ sectors, to define physical
off-shell correlation functions of gauge invariant quark and gluon fields,
without affecting the $S$ matrix amplitudes. The extension is made of
certain cloud bosons, plus their anticommuting partners. Both are quantized
as purely virtual, to ensure that they do not propagate ghosts. The extended
theory is renormalizable and unitary. In particular, the off-shell,
diagrammatic version of the optical theorem holds. We calculate the one-loop
two-point functions of dressed quarks and gluons, and show that their
absorptive parts are gauge independent, cloud independent and positive
(while they are generically unphysical if the cloud sectors are not purely
virtual). A gauge/cloud duality simplifies the computations and shows that
the gauge choice is just a particular cloud. It is possible to dress every
field insertion with a different cloud. We compare the purely virtual
extension to previous approaches to similar problems.

\vfill\eject

\section{Introduction}

\label{intro}\setcounter{equation}{0}

The greatest success of perturbative quantum field theory relies on the
theory of scattering. However, quantum field theory is not just scattering
processes. It is also off-shell correlation functions, including correlation
functions of composite fields. Gauge invariant composite fields can be
divided in two classes: those that are at least quadratic in the elementary
fields, and those that contain linear terms. It is straightforward to build
representatives of the first class, not equally easy to build composite
fields of the second class. The latter are particularly important, because
they provide a complete basis of observables and can eventually be used to
replace the elementary fields altogether. In this paper we extend quantum
field theory in a way that overcomes this difficulty and preserves the
fundamental physics.

Specifically, we add purely virtual \textquotedblleft
cloud\textquotedblright\ sectors\ to the Yang-Mills action, built by means
of certain cloud fields and their anticommuting partners. The sectors are
arranged so as to satisfy certain \textquotedblleft cloud
symmetries\textquotedblright , which ensure that the scattering amplitudes
coincide with the usual ones, and the correlation functions of the ordinary
fields are also unaffected. Each field insertion in a correlation function
can be rendered gauge invariant by \textquotedblleft
dressing\textquotedblright\ it with an independent cloud. Each cloud is
specified by a cloud function and a cloud Feddeev-Popov determinant.

To ensure that the extended theory is unitary and propagates no additional
degrees of freedom, we quantize the clouds as purely virtual \cite%
{diagrammarMio}. This way, the correlation functions of the dressed fields
satisfy the off-shell, diagrammatic version of the optical theorem.
Moreover, we show that the extended theory is renormalizable and polynomial
in all the fields except for the basic cloud fields (which are
dimensionless) and their anticommuting partners.

The extension is perturbative, and the expansion in powers of the gauge
coupling $g$ coincides with the expansion in the number of loops.
Renormalizability is proved to all orders by means of an extended
Batalin-Vilkovisky formalism and its Zinn-Justin master equations.

Note that the dressed fields we build are invariant under infinitesimal
gauge transformations, but are not required to be invariant under global
gauge transformations. This is indeed the way out to have physical non
singlet states without violating unitarity.

If we wish, we can use the formalism developed here to downgrade the
elementary fields (which are not gauge invariant) to mere integration and
diagrammatic tools, and use the dressed fields (which are manifestly gauge
invariant) everywhere. This way, we know from the start that everything we
compute is manifestly gauge independent.

As said, the purely virtual nature of the clouds ensures that no unwanted
degrees of freedom are propagated. This opens the way to extract physical
information from the off-shell correlation functions of the elementary
fields in a systematic way. We illustrate the basic properties of the
formalism by calculating the one-loop two-point functions of the dressed
quarks and gluons, and showing that their absorptive parts are physical.
When the clouds are not purely virtual (which occurs, form example, if we
quantize the cloud sectors by means of the Feynman $i\epsilon $
prescription), the absorptive parts are generically unphysical.

A certain gauge/cloud duality, which is sometimes helpful to simplify the
computations, shows that the usual gauge choice is ultimately nothing but a
particular cloud, as long as the gauge trivial modes are rendered purely
virtual. This suggests to use a \textquotedblleft purely virtual
gauge\textquotedblright\ as a valid alternative to the so-called physical
gauges \cite{physga}. Among those, we mention the Coulomb gauge, the
temporal gauge, the light-cone gauge and, more generally, the axial gauges.
Normally, such gauges lead to mathematical complications. What they miss is
the concept of pure virtuality, although in some cases (like the Coulomb
gauge), they incorporate it by accident, so to speak. In our approach, we do
not change the gauge fixing to make it physical. Rather, we make a
gauge-fixing physical by changing the prescription we use to define it.

We compare our formalism with other approaches to similar issues available
in\ the literature, putting particular emphasis on the \textquotedblleft
Coulomb\textquotedblright\ approaches, that is to say,\ the clouds defined
by Dirac in QED \cite{Dirac} and those studied by Lavelle and McMullan in
non-Abelian gauge theories \cite{Lavelle}. Earlier definitions of gauge
invariant variables in Yang-Mills theory are due to Chang \cite{Chang}.
Different lines of thinking exist as well, such as the 't Hooft approach,
based on composite fields and a symmetry breaking mechanism \cite%
{tHooftcloud}, and the approach based on Wilson lines.

In a parallel paper \cite{AbsoPhysGrav}, we explore similar issues in
gravity.

\bigskip

We point out some physical applications of our results. The main one is the
possibility of studying new types of scattering processes. As said, the
usual $S$ matrix amplitudes do not change, after the extension. Those
amplitudes concern asymptotic states, which become free in the infinite past
and in the infinite future. The correlation functions of dressed fields
overcome this restriction, and allow us to define \textquotedblleft
short-distance scattering processes\textquotedblright\ among colored states
of quarks and gluons, which are the processes where the incoming and
outgoing states are not allowed (or not not have enough time) to become
free. In the same spirit, we can study transition amplitudes in Yang-Mills
theories on compact manifolds, merging the formalism of this paper with the
one of \cite{MQ}, for situations where the experimental apparatus
surrounding the physical process actively influences the process itself. See
also \cite{MQQG} about this.

Experimental situations of this type are, among others, the interactions
inside a quark gluon plasma, or the interactions between quarks and gluons
at distances comparable to the proton radius, where we cannot use the notion
of asymptotic state. The increasing precision of present colliders and the
colliders of new generations make us hope that in a non dinstant future we
can be less dependent on the paradigms of quantum field theory that have
dominated the scene since its birth. The formalism of this paper breaks the
main technical barriers for the undertaking of such studies, and is a first
step towards devising feasable experiments.

The processes we have just mentioned are not, or do not need to be, on the
mass shell. If treated with the usual approaches, they are gauge dependent,
and unphysical. Our results imply that we can actually define them by means
of dressed fields, as long as the dressings are purely virtual. The results
we obtain are physical (i.e., gauge invariant and gauge independent -- in
addition, they obey the optical theorem), but depend on the dressing
parameters, which we denote by $\tilde{\lambda}$. These parameters do not
belong to the fundamental theory, but describe features of the experimental
setup, such as experimental resolutions, finite volume effects, finite
temperature effects, dependences on a background, or an external field, etc.

The $\tilde{\lambda}$ dependence of the results is not unexpected. Think,
for example, of the correlation functions built by means of Wilson lines:
they depend on the Wilson lines themselves. Another situation where the
physical predictions depend on the details of the instrumentation is when
the amplitudes are affected by infrared divergences, which are compensated
by soft and collinear photons, gluons, or gravitons \cite{infred}. In those
cases, the predictions depend on the energy resolution and the angular
resolution. Something similar occurs, to some extent, when we observe
unstable particles, like the muon \cite{muon}, which do not admit asymptotic
states in a strict sense.

The $\tilde{\lambda}$ dependencies mean that it is impossible to eliminate
the influence of the observer on the observed phenomenon. Yet, this does not
prevent us from making testable predictions. We can eliminate the $\tilde{%
\lambda}$ dependences by calibrating the instrumentation, i.e., by
sacrificing a few initial measurements to determine the values of the
parameters $\tilde{\lambda}$, after which everything is predicted uniquely,
and can be confirmed or falsified experimentally.

\bigskip

It is also useful to point out the differences between the goals of our
approach and the goals of other approaches to gauge theories that are
available in the literature, such as the compensator field approach \cite%
{compensator} and the Stueckelberg approach \cite{stueckelberg}. The first
one is a rephrasing of the theory and its gauge symmetries, but does not
change the cohomology of physical observables. The second one is used to
describe massive vectors. Our purpose, instead, is to define
\textquotedblleft gauge-invariant gauge fields\textquotedblright , so to
speak, that is to say, colored physical states of quarks and gluons. This is
possible by means of dressed fields.

The first difference between our approach and the compensator field approach
is that, after introducing the extra sectors, we still define the physical
observables as being gauge invariant: they are not required to be invariant
under the extra (cloud) transformations. The dressed fields, which are
indeed cloud dependent, are built on this premise. Once we have done that,
we can consider new correlation functions (those that contain insertions of
dressed fields) and study new scattering processes (the short distance
processes mentioned above). These goals cannot be achieved in the
compensator field approach. The correlation functions of ordinary gauge
invariant composite fields (built without using the could sector), such as $%
\bar{\psi}\psi $, $F_{\mu \nu }F^{\mu \nu }$, etc., and well as the $S$
matrix amplitudes, instead, do not change.

Since the dressed fields are just required to be gauge invariant, but not
cloud invariant, the extra fields become propagating. Generically, this can
be dangerous: if those fields are not treated properly, they may affect the
observable spectra in undesirable ways. We show, by means of explicit
calculations, that if they are quantized by means of the usual Feynman
prescription, they inject ghosts into the theory. Since our definition of
physical fields prevents us from getting rid of them cohomologically, we
must achieve the goal in a different, non cohomological way: we make them
purely virtual.

To make the whole construction work, we need to keep the usual sector and
the cloud sectors to some extent separated. In particular, the functions
that define the clouds should be gauge invariant, while the usual
gauge-fixing functions should be cloud invariant. We show that these
restrictions are consistent, because they are preserved by renormalization.
Restrictions on the gauge-fixing choices are not unusual. A familiar one is
adopted in the context of the background field method, where the
gauge-fixing must be invariant under the background transformations.

\bigskip

We recall that purely virtual particles, also called fake particles, or
\textquotedblleft fakeons\textquotedblright , are defined by a new
diagrammatics \cite{diagrammarMio}, which takes advantage of the possibility
of splitting the usual optical theorem \cite{unitarity} into independent,
algebraic spectral optical identities. Each identity is associated with a
different (multi)threshold. By removing subsets of such identities, and
projecting the whole theory to the physical subspace, certain degrees of
freedom can be removed at all energies, while preserving unitarity and the
optical theorem in a manifest way. The main application of this idea is the
formulation of a consistent theory of quantum gravity \cite{LWgrav}, which
is observationally testable due to its predictions in inflationary cosmology 
\cite{ABP}. At the phenomenological level, fakeons evade common constraints
that limit the employment of normal particles \cite{Tallinn1,Tallinn2}.

Throughout the paper we work with the dimensional regularization \cite%
{dimreg}, $\varepsilon =4-D$ denoting the difference between the physical
dimension and the continued one.

The paper is organized as follows. In section \ref{cloudfields} we give the
basic definitions that are necessary to build the cloud sectors. In section %
\ref{Bata} we recall the standard Batalin-Vilkovisky formalism for gauge
theories, and the Zinn-Justin master equation. In section \ref{BataCloud} we
extend the formalism and the master equation to define the cloud sector. In
section \ref{cloudindep} we show that the ordinary correlation functions of
elementary and composite fields are unaffected by the cloud sector. In
section \ref{cloudindepS} we prove the same for the $S$ matrix amplitudes.
In section \ref{cloudcorre} we build the correlation functions of the
dressed fields. In section \ref{dual} we prove that the cloud sector and the
gauge-trivial sector are related by a certain duality relation. In section %
\ref{multiclouds} we add several copies of the could sector and show that
each insertion in a correlation function can be dressed with its own,
independent cloud. In section \ref{absorpt} we define the absorptive parts
and study their properties. In section \ref{drfermion} we compute the
two-point function of the dressed fermions at one loop with a covariant
cloud and show that its absorptive part is unphysical. In section \ref{pvc}
we overcome this difficulty by introducing purely virtual clouds. In section %
\ref{gauge2} we repeat the analysis for the two-point function of the
dressed gauge fields, and show that the absorptive part is physical, if
purely virtual clouds are used. In section \ref{renormalization} we prove
that the extended theory is renormalizable, and show how the renormalization
works in detail. In section \ref{comparison} we compare our approach with
other approaches available in the literature. Section \ref{conclusions}
contains the conclusions, while appendix \ref{formulas} contains the
notation and some useful formulas. In appendix \ref{uniqueness} we prove
that the dressed fields are unique. In appendix \ref{appe3} we study how the
cloud independence goes through renormalization.

\section{The cloud field, its anticommuting partner, and the dressed fields}

\setcounter{equation}{0}\label{cloudfields}

In this section we lay out the basic notions that are needed to build the
cloud sectors. For definiteness, we consider Yang-Mills theory with gauge
group $G=SU(N_{c})$ and quarks $\psi $ in the fundamental representation.

The dressings can be easily worked out, once the theory contains a field $%
\hat{U}(x)$, with values in $G$, that transforms as%
\begin{equation}
\delta _{\Lambda }\hat{U}=-ig\hat{U}\Lambda  \label{dLU}
\end{equation}%
under a gauge transformation. Here, $\Lambda (x)=\Lambda ^{a}(x)T^{a}$ are
the parameters of the transformation and $T^{a}$ are the Hermitian matrices
of the fundamental representation.

For example, if $\psi $ is a fermion, the product $\hat{U}\psi $ is
obviously gauge invariant, because (\ref{dLU}) and the $\psi $
transformation law $\delta _{\Lambda }\psi =ig\Lambda \psi $ imply $\delta
_{\Lambda }(\hat{U}\psi )=0$. Nevertheless, it is not convenient to use $%
\hat{U}$ as an elementary field for the perturbative expansion, since we
also need the inverse matrix $\hat{U}^{-1}$. It is better to write $\hat{U}=%
\mathrm{e}^{-ig\phi }$, and define $\phi =\phi ^{a}T^{a}$ as the fundamental
\textquotedblleft cloud field\textquotedblright .

The gauge transformation $\delta _{\Lambda }\phi $ of $\phi $ can be derived
from (\ref{dLU}). A version of the Campbell-Baker-Hausdorff formula reads%
\begin{equation}
\mathrm{de}^{X}=\mathrm{e}^{X}\sum_{n=0}^{\infty }\frac{(-\mathrm{ad}%
_{X})^{n}}{(n+1)!}\mathrm{d}X=\mathrm{e}^{X}\frac{1-\mathrm{e}^{-\mathrm{ad}%
_{X}}}{\mathrm{ad}_{X}}\mathrm{d}X  \label{dex}
\end{equation}%
where $\mathrm{ad}_{X}Y\equiv \lbrack X,Y]$, $X$ and $Y$ being matrices or
operators. After rearranging the formula into the form%
\begin{equation*}
\mathrm{d}X=\frac{\mathrm{ad}_{X}}{1-\mathrm{e}^{-\mathrm{ad}_{X}}}\mathrm{e}%
^{-X}\mathrm{de}^{X},
\end{equation*}%
we apply it with $\hat{U}=\mathrm{e}^{-ig\phi }$, $X=-ig\phi $, $\mathrm{de}%
^{X}\rightarrow \delta _{\Lambda }\hat{U}$, $\mathrm{d}\phi \rightarrow
\delta _{\Lambda }\phi $. This way, the desired $\phi $ gauge transformation
is easily found. It reads%
\begin{eqnarray}
\delta _{\Lambda }\phi &=&\frac{ig\hspace{0.01in}\mathrm{ad}_{\phi }}{%
\mathrm{e}^{ig\hspace{0.01in}\mathrm{ad}_{\phi }}-1}\Lambda \equiv R(\phi
,\Lambda )\equiv T^{a}R^{a}(\phi ,\Lambda )  \notag \\
&=&\Lambda -\frac{ig}{2}[\phi ,\Lambda ]-\frac{g^{2}}{12}[\phi ,[\phi
,\Lambda ]]+\mathcal{O}(g^{3}).  \label{tfl}
\end{eqnarray}

For the moment, we define the cloud field $\phi $ as a field that transforms
according to this law. In the next sections we introduce it at the level of
the action, derive its Feynman rules and study its diagrammatic properties.

It is possible to check the closure of the transformation, i.e.,%
\begin{equation}
\lbrack \delta _{\Lambda },\delta _{\Sigma }]\phi =\delta _{-ig[\Lambda
,\Sigma ]}\phi ,  \label{clos}
\end{equation}%
where $-ig[\Lambda ,\Sigma ]=gT^{a}f^{abc}\Lambda ^{b}\Sigma ^{c}$.

The gauge-invariant dressed fields are%
\begin{eqnarray}
A_{\mu \text{d}} &=&\hat{U}A_{\mu }\hat{U}^{-1}+\frac{i}{g}\hat{U}(\partial
_{\mu }\hat{U}^{-1})=\mathrm{e}^{-ig\hspace{0.01in}\mathrm{ad}_{\phi
}}A_{\mu }-\frac{1-\mathrm{e}^{-ig\hspace{0.01in}\mathrm{ad}_{\phi }}}{ig%
\hspace{0.01in}\mathrm{ad}_{\phi }}(\partial _{\mu }\phi ),  \notag \\
\psi _{\text{d}} &=&\mathrm{e}^{-ig\phi }\psi ,\quad \bar{\psi}_{\text{d}}=%
\bar{\psi}\mathrm{e}^{ig\phi }.  \label{dressedfields}
\end{eqnarray}%
The explicit expression of $A_{\mu \text{d}}$ is obtained by means of (\ref%
{dex}) and $\mathrm{e}^{X}Y\mathrm{e}^{-X}=\mathrm{e}^{\mathrm{ad}_{X}}Y$.

It is easy to prove that the fields (\ref{dressedfields}) are indeed gauge
invariant: $\delta _{\Lambda }A_{\mu \text{d}}=\delta _{\Lambda }\psi _{%
\text{d}}=\delta _{\Lambda }\bar{\psi}_{\text{d}}=0$. It is also possible to
prove (see appendix \ref{uniqueness}) that they are unique, given the $\phi $
transformation law.

It is also crucial to introduce an anticommuting partner $H=H^{a}T^{a}$ of $%
\phi $, transforming as%
\begin{equation}
\delta _{\Lambda }H^{a}=H^{b}\frac{\delta R^{a}(\phi ,\Lambda )}{\delta \phi
^{b}}.  \label{dH}
\end{equation}%
The consistency of this transformation law can be readily proved from its
closure:%
\begin{eqnarray}
\lbrack \delta _{\Lambda },\delta _{\Sigma }]H^{a} &=&H^{b}\frac{\delta
R^{c}(\phi ,\Lambda )}{\delta \phi ^{b}}\frac{\delta R^{a}(\phi ,\Sigma )}{%
\delta \phi ^{c}}+H^{b}R^{c}(\phi ,\Lambda )\frac{\delta ^{2}R^{a}(\phi
,\Sigma )}{\delta \phi ^{b}\delta \phi ^{c}}-(\Lambda \leftrightarrow \Sigma
)  \notag \\
&=&H^{b}\frac{\delta }{\delta \phi ^{b}}\left( R^{c}(\phi ,\Lambda )\frac{%
\delta R^{a}(\phi ,\Sigma )}{\delta \phi ^{c}}-R^{c}(\phi ,\Sigma )\frac{%
\delta R^{a}(\phi ,\Lambda )}{\delta \phi ^{c}}\right)  \notag \\
&=&H^{b}\frac{\delta R^{a}(\phi ,-ig[\Lambda ,\Sigma ])}{\delta \phi ^{b}}%
=\delta _{-ig[\Lambda ,\Sigma ]}H^{a}.\qquad \,  \label{ddH}
\end{eqnarray}%
We have used (\ref{dH}) in the first line, (\ref{clos}) in the second line
and (\ref{dH}) again in the last step.

We have achieved what we wanted, that is to say, define gauge-invariant
dressings for quarks and gluons. However, we have done it at the cost of
introducing new fields: the cloud field $\phi $ and its anticommuting
partner $H$. The next problem is to include the extra fields into the
action, and ensure that:

$a$) the extension does not change the fundamental physics;

$b$) in particular, no unphysical degrees of freedom propagate.

The field $H$ plays a crucial role to achieve objective $a$). Specifically,
we use it to endow the cloud sector with a certain symmetry, which ensures
that the correlation functions of the undressed fields are unmodified
(despite the presence of nontrivial interactions between them and the extra
fields), and so are the $S$ matrix amplitudes. Moreover, we render the extra
fields purely virtual, to ensure that requirement $b$) is manifestly
fulfilled as well.

We also want to preserve locality, renormalizability and unitarity, and do
everything without affecting the usual structure of the perturbative
expansion. In particular, the expansion in powers of the gauge coupling $g$
should coincide with the expansion in the number of loops.

\section{Batalin-Vilkovisky formalism and Zinn-Justin master equation}

\setcounter{equation}{0}\label{Bata}

In this section we recall the standard formalism to treat gauge theories. In
the next sections we generalize it to build the cloud sector.

We start from the classical action%
\begin{equation}
S_{\text{cl}}(A,\bar{\psi},\psi )=-\frac{1}{4}\int F_{\mu \nu }^{a}F^{\mu
\nu a}+\int \bar{\psi}(i\gamma ^{\mu }D_{\mu }-m)\psi  \label{slc}
\end{equation}%
of a non-Abelian gauge theory with gauge group $SU(N_{c})$, coupled to
matter. For concreteness, we assume that the matter sector is made of
fermions $\psi $ in the fundamental representation, $D_{\mu }\psi =\partial
_{\mu }\psi -igA_{\mu }\psi $ being their covariant derivative. The specific
form of $S_{\text{cl}}$ is not important for the formalism we are going to
develop. However, (\ref{slc}) will be used in the explicit computations of
this paper. We do not write the measure \textrm{d}$^{D}x$ of the spacetime
integrals explicitly, when no confusion can arise.

We introduce the set of fields $\Phi ^{\alpha }=(A_{\mu },C,\bar{C},B,\psi ,%
\bar{\psi})$, where $A_{\mu }={T^{a}A_{\mu }^{a}}$ are the gauge fields, $%
C=C^{a}T^{a}$ are the Faddeev-Popov ghosts \cite{FP}, $B=B^{a}T^{a}$ are the
Nakanishi-Lautrup Lagrange multipliers \cite{naka} and $\bar{C}=\bar{C}%
^{a}T^{a}$ are the antighosts. The superscript $\alpha $ collects all the
indices. To have control on the Ward-Takahashi-Slavnov-Taylor identities 
\cite{WTST} to all orders in a compact form, we use the Batalin-Vilkovisky
formalism \cite{BV}.

We couple sources $K^{\alpha }=(K_{A}^{\mu },K_{C},K_{\bar{C}},K_{B},{%
K_{\psi },K_{\bar{\psi}}})$ to the field transformations by means of the
functional%
\begin{equation}
S_{K}(\Phi ,K)=-\int (D_{\mu }C)^{a}K_{A}^{\mu a}+\frac{g}{2}\int
f^{abc}C^{b}C^{c}K_{C}^{a}-i{g\int \bar{\psi}CK_{\bar{\psi}}-ig\int K_{\psi
}C\psi }-\int B^{a}K_{\bar{C}}^{a},  \label{SK}
\end{equation}%
where $D_{\mu }C^{a}=\partial _{\mu }C^{a}+gf^{abc}A_{\mu }^{b}C^{c}$ is the
covariant derivative of $C^{a}$. Precisely, the infinitesimal field
transformations are%
\begin{equation}
\delta _{\Lambda }\Phi ^{\alpha }=\theta (S_{K},\Phi ^{\alpha })=-\theta 
\frac{\delta _{r}S_{K}}{\delta K^{\alpha }},  \label{gaugetr}
\end{equation}%
where $\Lambda =\theta C$, $\theta $ is a constant anticommuting (Grassmann)
variable and%
\begin{equation}
(X,Y)=\int \left( \frac{\delta _{r}X}{\delta \Phi ^{\alpha }}\frac{\delta
_{l}Y}{\delta K^{\alpha }}-\frac{\delta _{r}X}{\delta K^{\alpha }}\frac{%
\delta _{l}Y}{\delta \Phi ^{\alpha }}\right)  \label{brackets}
\end{equation}%
are the Batalin-Vilkovisky antiparentheses \cite{BV}, the subscripts $r$ and 
$l$ denoting the right and left derivatives, respectively.

The closure of the algebra of transformations is encoded into the identities 
\begin{equation}
(S_{K},S_{K})=0,\qquad (S_{K},(S_{K},X))=0  \label{closure}
\end{equation}%
for every $X$. The Jacobi identity satisfied by the antiparentheses \cite{BV}
implies that the two properties just stated are equivalent. The second one
is called nilpotence relation.

The gauge-fixed action reads%
\begin{equation}
S_{\text{gf}}(\Phi )=S_{\text{cl}}(A,\bar{\psi},\psi )+(S_{K},\Psi (\Phi )),
\label{Sgf}
\end{equation}%
where $\Psi (\Phi )$ is a certain functional that fixes the gauge, commonly
known as \textquotedblleft gauge fermion\textquotedblright . A typical form
of $\Psi (\Phi )$ is 
\begin{equation}
\Psi (\Phi )=\int \bar{C}^{a}\left( G^{a}(A)+\frac{\lambda }{2}B^{a}\right) ,
\label{gferm}
\end{equation}%
where $G^{a}(A)$ is the gauge-fixing function. For example, the covariant
gauge is the one with $G^{a}(A)=\partial ^{\mu }A_{\mu }^{a}$, which gives%
\begin{equation}
(S_{K},\Psi )=\int B^{a}\left( \partial ^{\mu }A_{\mu }^{a}+\frac{\lambda }{2%
}B^{a}\right) -\int \bar{C}^{a}\partial ^{\mu }D_{\mu }C^{a}\rightarrow -%
\frac{1}{2\lambda }\int \left( \partial ^{\mu }A_{\mu }^{a}\right) ^{2}-\int 
\bar{C}^{a}\partial ^{\mu }D_{\mu }C^{a},  \label{covg}
\end{equation}%
where the arrow denotes the integration over $B^{a}$. The gauge-fixed action
then reads%
\begin{equation}
S_{\text{gf}}(\Phi )=-\frac{1}{4}\int F_{\mu \nu }^{a}F^{\mu \nu a}+\int 
\bar{\psi}(i\gamma ^{\mu }D_{\mu }-m)\psi -\frac{1}{2\lambda }\int \left(
\partial ^{\mu }A_{\mu }^{a}\right) ^{2}-\int \bar{C}^{a}\partial ^{\mu
}D_{\mu }C^{a}.  \label{sgfconcr}
\end{equation}%
Other gauge choices will be considered in the paper.

To have control on the renormalization of the gauge transformations, it is
useful to include them as composite fields. This is achieved by adding $%
S_{K} $ to the gauge-fixed action and working with the new action%
\begin{equation}
S(\Phi ,K)=S_{\text{gf}}(\Phi )+S_{K}(\Phi ,K).  \label{action}
\end{equation}%
The identities (\ref{closure}) imply that $S(\Phi ,K)$ satisfies the
Zinn-Justin equation \cite{ZJ} 
\begin{equation}
(S,S)=0,  \label{master}
\end{equation}%
also known as master equation. Formula (\ref{master}) collects the gauge
invariance of the classical action, the triviality of the gauge-fixing
sector, and the closure of the gauge algebra. The Jacobi identity implies
the nilpotence relation $(S,(S,X))=0$ for every $X$.

\section{Cloud sector}

\setcounter{equation}{0}\label{BataCloud}

In this section we define the cloud sector. The idea is to add the cloud
field $\phi $ to the action, but trivialize its presence, in some sense, by
means of a new symmetry (which we call cloud symmetry), built with the
anticommuting partner $H$, so as to keep the correlation functions built
without involving the cloud sector and the $S$ matrix elements unchanged.

The goal is achieved as follows. First, we introduce a new set of fields $%
\tilde{\Phi}^{\alpha }=(\phi ^{a},H^{a},\bar{H}^{a},$ $E^{a})$ and the
sources $\tilde{K}^{\alpha }=(\tilde{K}_{\phi }^{a},\tilde{K}_{H}^{a},\tilde{%
K}_{\bar{H}}^{a},\tilde{K}_{E}^{a})$ coupled to their transformations, where 
$H$ can be understood as \textquotedblleft cloud Faddeev-Popov
ghosts\textquotedblright , $E$ are new Lagrange multipliers and $\bar{H}$
are the cloud antighosts. Second, we extend the definition (\ref{brackets})
of antiparentheses to include the new sector:%
\begin{equation*}
(X,Y)=\int \left( \frac{\delta _{r}X}{\delta \Phi ^{\alpha }}\frac{\delta
_{l}Y}{\delta K^{\alpha }}-\frac{\delta _{r}X}{\delta K^{\alpha }}\frac{%
\delta _{l}Y}{\delta \Phi ^{\alpha }}+\frac{\delta _{r}X}{\delta \tilde{\Phi}%
^{\alpha }}\frac{\delta _{l}Y}{\delta \tilde{K}^{\alpha }}-\frac{\delta _{r}X%
}{\delta \tilde{K}^{\alpha }}\frac{\delta _{l}Y}{\delta \tilde{\Phi}^{\alpha
}}\right) .
\end{equation*}%
Third, we collect the transformations of the old and new fields into the
functionals 
\begin{eqnarray}
S_{K}^{\text{gauge}} &=&S_{K}-\int R^{a}(\phi ,C)\tilde{K}_{\phi }^{a}+\int
H^{b}\frac{\delta R^{a}(\phi ,C)}{\delta \phi ^{b}}\hspace{0.01in}\tilde{K}%
_{H}^{a},  \notag \\
S_{K}^{\text{cloud}} &=&\int H^{a}\tilde{K}_{\phi }^{a}-\int E^{a}\tilde{K}_{%
\bar{H}}^{a},\qquad S_{K}^{\text{tot}}=S_{K}^{\text{gauge}}+S_{K}^{\text{%
cloud}},  \label{sk}
\end{eqnarray}%
where $R^{a}(\phi ,C)$, defined in (\ref{tfl}), is just another way to write 
$\delta _{C}\phi ^{a}$. The first functional collects the gauge
transformations (\ref{tfl})\ and (\ref{dH}) of $\phi $ and $H$, while the
second functional encodes the cloud transformations, which are the most
general shifts of $\phi $ and $\bar{H}$. For example, the total
transformation of $\phi $ is 
\begin{equation*}
\delta _{\Lambda ,\mathcal{H}}\phi ^{a}=\theta \left( S_{K}^{\text{tot}%
},\phi ^{a}\right) =-\theta H^{a}+R^{a}(\phi ,\Lambda )=-\mathcal{H}%
^{a}+\delta _{\Lambda }\phi ^{a},
\end{equation*}%
where $\mathcal{H}=\theta H$ can be viewed as an arbitrary function that
translates the cloud field $\phi $.

It is easy to check the identities%
\begin{equation}
(S_{K}^{\text{gauge}},S_{K}^{\text{gauge}})=(S_{K}^{\text{cloud}},S_{K}^{%
\text{cloud}})=0,  \label{ss1}
\end{equation}%
which express the closures of the algebras of the gauge and cloud
transformations. The first identity follows from (\ref{closure}), (\ref{clos}%
) and (\ref{ddH}).

We can also check the closure of the combined transformations, i.e.,%
\begin{equation}
(S_{K}^{\text{tot}},S_{K}^{\text{tot}})=(S_{K}^{\text{cloud}},S_{K}^{\text{%
gauge}})=0.  \label{ss2}
\end{equation}%
The proof follows from 
\begin{equation*}
(S_{K}^{\text{cloud}},S_{K}^{\text{gauge}})=\int \left( \frac{\delta
_{r}S_{K}^{\text{cloud}}}{\delta H^{a}}\frac{\delta _{l}S_{K}^{\text{gauge}}%
}{\delta \tilde{K}_{H}^{a}}-\frac{\delta _{r}S_{K}^{\text{cloud}}}{\delta 
\tilde{K}_{\phi }^{a}}\frac{\delta _{l}S_{K}^{\text{gauge}}}{\delta \phi ^{a}%
}\right) =0.
\end{equation*}

The following identities also hold:%
\begin{equation}
S_{K}^{\text{gauge}}=S_{K}-\left( S_{K}^{\text{cloud}},\int R^{a}(\phi ,C)%
\tilde{K}_{H}^{a}\right) ,\qquad S_{K}^{\text{cloud}}=-\left( S_{K}^{\text{%
cloud}},\int \phi ^{a}\tilde{K}_{\phi }^{a}+\int \bar{H}^{a}\tilde{K}_{\bar{H%
}}^{a}\right) .  \label{exa}
\end{equation}%
They show that the two functionals $S_{K}^{\text{gauge}}-S_{K}$ and $S_{K}^{%
\text{cloud}}$ are \textquotedblleft cohomologically
exact\textquotedblright\ under the cloud symmetry, i.e., they have the form $%
\left( S_{K}^{\text{cloud}},\text{local functional}\right) $.

\subsection{The cloud and the total action}

To specify the cloud we want to use, we add $(S_{K}^{\text{tot}},\tilde{\Psi}%
)$ to the action, where $\tilde{\Psi}(\Phi ,\tilde{\Phi})$ is the
\textquotedblleft cloud fermion\textquotedblright . A\ typical form of it
is\ 
\begin{equation}
\tilde{\Psi}(\Phi ,\tilde{\Phi})=\int \bar{H}^{a}\left( V^{a}(\phi ,A)+\frac{%
\tilde{\lambda}}{2}E^{a}\right) ,  \label{cloud fermion}
\end{equation}%
where $V^{a}(\phi ,A)$ denotes the \textquotedblleft cloud-fixing
function\textquotedblright , i.e., the function that defines the cloud. We
assume that $V^{a}$ is gauge invariant, 
\begin{equation}
(S_{K}^{\text{gauge}},V^{a})=0,\qquad (S_{K}^{\text{gauge}},\tilde{\Psi})=0.
\label{spsit}
\end{equation}%
In practice, $V^{a}(\phi ,A)$ depends on $\phi $ and $A_{\mu }$ only through
the dressed gauge field $A_{\mu \text{d}}$. Sometimes, with an abuse of
notation, we just write $V^{a}(A_{\mu \text{d}})$. More generally, $V^{a}$
may depend on the other dressed fields as well.

The gauge fermion (\ref{gferm}) was implicitly assumed to be cloud
invariant, because it was built before adding the extra sectors. This is an
assumption we have to maintain after the extension, and will be crucial for
the construction of the correlation functions of the dressed fields. In the
end, the cloud-fermion must be gauge invariant and the gauge-fermion must be
cloud invariant.

We find 
\begin{equation}
(S_{K}^{\text{tot}},\tilde{\Psi})=(S_{K}^{\text{cloud}},\tilde{\Psi})=\int
E^{a}\left( V^{a}(\phi ,A)+\frac{\tilde{\lambda}}{2}E^{a}\right) +\int \bar{H%
}^{a}\frac{\delta V^{a}(\phi ,A)}{\delta \phi ^{b}}H^{b}.  \label{scloud}
\end{equation}%
Note that the last term provides a sort of Faddeev-Popov determinant for the
cloud, which is crucial for the properties that we want to prove.

The total action of the extended theory is then 
\begin{equation}
S_{\text{tot}}(\Phi ,K,\tilde{\Phi},\tilde{K})=S_{\text{cl}}+(S_{K}^{\text{%
tot}},\Psi +\tilde{\Psi})+S_{K}^{\text{tot}}  \label{stota}
\end{equation}%
and satisfies its own master equation 
\begin{equation}
(S_{\text{tot}},S_{\text{tot}})=0.  \label{cloudmaster}
\end{equation}%
Note that $S_{\text{tot}}$ is separately gauge invariant and cloud
invariant, since equations (\ref{ss1}), (\ref{ss2}) and (\ref{spsit}) imply%
\begin{equation}
(S_{K}^{\text{gauge}},S_{\text{tot}})=(S_{K}^{\text{cloud}},S_{\text{tot}%
})=0.  \label{stot}
\end{equation}

Finally, using (\ref{exa}) we can write%
\begin{equation*}
S_{\text{tot}}(\Phi ,K,\tilde{\Phi},\tilde{K})=S(\Phi ,K)+(S_{K}^{\text{cloud%
}},\Theta ),\qquad \Theta =\tilde{\Psi}-\int R^{a}(\phi ,C)\tilde{K}%
_{H}^{a}-\int \phi ^{a}\tilde{K}_{\phi }^{a}-\int \bar{H}^{a}\tilde{K}_{\bar{%
H}}^{a},
\end{equation*}%
which shows that the difference between the total action and the ordinary
action $S(\Phi ,K)$ of formula (\ref{action}) is exact under the cloud
symmetry.

\subsection{Covariant cloud and propagators}

To make explicit calculations, we need to choose the could function. There
is a large arbitrariness in this choice. A convenient starting point is the
covariant cloud 
\begin{equation}
V^{a}(\phi ,A)=\partial ^{\mu }A_{\mu \text{d}}^{a}.  \label{cloud}
\end{equation}%
Other choices will be considered later on.

Formula (\ref{scloud}) then gives%
\begin{equation}
(S_{K}^{\text{tot}},\tilde{\Psi})=(S_{K}^{\text{cloud}},\tilde{\Psi})=-\frac{%
1}{2\tilde{\lambda}}\int \left( \partial ^{\mu }A_{\mu \text{d}}^{a}\right)
^{2}-\int (\partial ^{\mu }\bar{H}^{b})\frac{\delta A_{\mu \text{d}}^{b}}{%
\delta \phi ^{a}}H_{\text{d}}^{a},  \label{sKpsit}
\end{equation}%
after integrating $E$ out. To the lowest order, (\ref{sKpsit}) reads%
\begin{equation}
(S_{K}^{\text{cloud}},\tilde{\Psi})=-\frac{1}{2\tilde{\lambda}}\int \left(
\partial ^{\mu }A_{\mu }^{a}-\square \phi ^{a}\right) ^{2}+\int (\partial
^{\mu }\bar{H}^{a})(\partial _{\mu }H^{a})+\mathcal{O}(g).  \label{cloudkin}
\end{equation}%
Together with (\ref{sgfconcr}), this expression allows us to derive the
propagators in the covariant framework. We find%
\begin{eqnarray}
\langle A_{\mu }^{a}(p)\hspace{0.01in}A_{\nu }^{b}(-p)\rangle _{0} &=&-\frac{%
i\delta ^{ab}}{p^{2}+i\epsilon }\left( \eta _{\mu \nu }-\frac{(1-\lambda
)p_{\mu }p_{\nu }}{p^{2}+i\epsilon }\right) ,\qquad \langle \psi (p)\hspace{%
0.01in}\bar{\psi}(-p)\rangle _{0}=\frac{i(\gamma ^{\mu }p_{\mu }+m)}{%
p^{2}-m^{2}+i\epsilon },  \notag \\
\langle A_{\mu }^{a}(p)\hspace{0.01in}\phi ^{b}(-p)\rangle _{0} &=&-\frac{%
\lambda \delta ^{ab}p_{\mu }}{(p^{2}+i\epsilon )^{2}},\qquad \langle \phi
^{a}(p)\hspace{0.01in}\phi ^{b}(-p)\rangle _{0}=-\frac{i(\lambda +\tilde{%
\lambda})\delta ^{ab}}{(p^{2}+i\epsilon )^{2}},  \label{propa}
\end{eqnarray}%
plus the ghost propagators. With an abuse of notation, we use the same
symbols for the fields and their Fourier transforms, since the meaning is
clear from the context. For the moment, we use the Feynman $i\epsilon $
prescription for every pole. Later on we switch to the purely virtual
(fakeon) prescription for the cloud poles.

The correlation functions of the dressed fields do not suffer from infrared
divergences (even without advocating the properties of the dimensional
regularization), although the denominators of some propagators contain the
square of $p^{2}$. A quick way to prove this statement is by noting that
there exists a gauge choice ($\lambda =-\tilde{\lambda}$) where such
problems are manifestly absent. Indeed, the concerning denominators are only
those originated by $\langle \phi ^{a}(p)\hspace{0.01in}\phi ^{b}(-p)\rangle
_{0}$, which vanishes for $\lambda =-\tilde{\lambda}$. Since the dressed
correlation functions are gauge independent, they are also infrared finite.
With more general gauge choices (such as $\lambda \neq -\tilde{\lambda}$),
the infrared divergences cancel out among different diagrams contributing to
the same order.

\section{Cloud independence of the ordinary correlation functions}

\setcounter{equation}{0}\label{cloudindep}

In this section and the next one we prove the cloud independence of the
non-cloud sector. We start by showing that the ordinary correlation
functions are unaffected by the clouds. This also ensures that the
renormalization of the fundamental (i.e., non-cloud) sector of the theory is
the same as usual.

The generating functional of the correlation functions is%
\begin{equation}
Z_{\text{tot}}(J,K,\tilde{J},\tilde{K})=\int [\mathrm{d}\Phi \mathrm{d}%
\tilde{\Phi}]\mathrm{\exp }\left( iS_{\text{tot}}(\Phi ,K,\tilde{\Phi},%
\tilde{K})+i\int \Phi ^{\alpha }J^{\alpha }+i\int \tilde{\Phi}^{\alpha }%
\tilde{J}^{\alpha }\right)  \label{Z}
\end{equation}%
and $W_{\text{tot}}(J,K,\tilde{J},\tilde{K})=-i\ln Z_{\text{tot}}(J,K,\tilde{%
J},\tilde{K})$ is the generating functional of the connected ones. The
functional derivatives of $Z_{\text{tot}}$ or $W_{\text{tot}}$ with respect
to the sources $J^{\alpha }$, calculated at $\tilde{J}=\tilde{K}=0$, are the
ordinary correlation functions of the elementary fields (and their
transformations). They are collected in 
\begin{eqnarray}
Z_{\text{tot}}(J,K,0,0) &=&\int [\mathrm{d}\Phi \mathrm{d}\tilde{\Phi}]%
\mathrm{\exp }\left( iS(\Phi ,K)+i(S_{K}^{\text{cloud}},\tilde{\Psi})+i\int
\Phi ^{\alpha }J^{\alpha }\right)  \notag \\
&=&\int [\mathrm{d}\Phi ]\mathrm{\exp }\left( iS(\Phi ,K)+i\int \Phi
^{\alpha }J_{\alpha }\right) \int [\mathrm{d}\tilde{\Phi}]\mathrm{e}%
^{i(S_{K}^{\text{cloud}},\tilde{\Psi})}.  \label{Z0}
\end{eqnarray}%
We want to prove that this expression coincides with the ordinary generating
functional, thanks to the identity%
\begin{equation}
\int [\mathrm{d}\tilde{\Phi}]\mathrm{e}^{i(S_{K}^{\text{cloud}},\tilde{\Psi}%
)}=1.  \label{ide}
\end{equation}

Using (\ref{scloud}) and integrating over the cloud ghosts and antighosts,
the left-hand side of (\ref{ide}) becomes%
\begin{equation*}
\int [\mathrm{d}\phi \mathrm{d}E]\mathrm{\exp }\left( i\int E^{a}\left(
V^{a}+\frac{\tilde{\lambda}}{2}E^{a}\right) \right) \det \left[ \frac{\delta
V^{a}}{\delta \phi ^{b}}\right] .
\end{equation*}%
Inserting%
\begin{equation*}
1=\int [\mathrm{d}Q]\mathrm{\exp }\left( -\frac{i\tilde{\lambda}}{2}\int
(Q^{a}-E^{a})^{2}\right) ,
\end{equation*}%
it also becomes%
\begin{equation*}
\int [\mathrm{d}\phi \mathrm{d}Q\mathrm{d}E]\mathrm{e}^{-\frac{i\tilde{%
\lambda}}{2}\int (Q^{a})^{2}}\mathrm{\exp }\left( i\int E^{a}(V^{a}+\tilde{%
\lambda}Q^{a})\right) \det \left[ \frac{\delta V^{a}}{\delta \phi ^{b}}%
\right] .
\end{equation*}%
Integrating on the Lagrange multipliers $E$, we obtain a functional $\delta $
function, and conclude 
\begin{equation*}
\int [\mathrm{d}Q]\mathrm{e}^{-\frac{i\tilde{\lambda}}{2}\int
(Q^{a})^{2}}\int [\mathrm{d}\phi ]\delta (V^{a}+\tilde{\lambda}Q^{a})\det %
\left[ \frac{\delta V^{a}}{\delta \phi ^{b}}\right] =\int [\mathrm{d}Q]%
\mathrm{e}^{-\frac{i\tilde{\lambda}}{2}\int (Q^{a})^{2}}=1,
\end{equation*}%
as desired.

Note that the cloud Faddeev-Popov determinant\ is crucial to trivialize the $%
\phi $ integral. Without it, the cloud sector would affect the non-cloud one
and change the fundamental theory.

\section{Cloud independence of the \emph{S} matrix amplitudes}

\setcounter{equation}{0}\label{cloudindepS}

Now we prove that the scattering amplitudes of the dressed fields coincide
with the usual scattering amplitudes (of undressed fields). Specifically,
the clouds have no effect on shell, when the polarizations are attached to
the amputated external legs.

First, consider a generic theory of scalar fields $\varphi $, described by
some classical action $S(\varphi )$. If $\mathcal{O}(\varphi )$ denotes a
composite field that is at least quadratic in $\varphi $, the connected
two-point function of $\varphi ^{\prime }\equiv \varphi +\mathcal{O}(\varphi
)$ can be decomposed, in momentum space, as%
\begin{eqnarray}
\langle \varphi ^{\prime }\hspace{0.01in}|\hspace{0.01in}\varphi ^{\prime
}\rangle &=&\langle \varphi \hspace{0.01in}|\hspace{0.01in}\varphi \rangle +%
\llangle\mathcal{O}\hspace{0.01in}|\hspace{0.01in}\varphi \rrangle\langle
\varphi \hspace{0.01in}|\hspace{0.01in}\varphi \rangle +\langle \varphi 
\hspace{0.01in}|\hspace{0.01in}\varphi \rangle \llangle\varphi \hspace{0.01in%
}|\hspace{0.01in}\mathcal{O}\rrangle+\llangle\mathcal{O}\hspace{0.01in}|%
\hspace{0.01in}\mathcal{O}\rrangle+\llangle\mathcal{O}\hspace{0.01in}|%
\hspace{0.01in}\varphi \rrangle\langle \varphi \hspace{0.01in}|\hspace{0.01in%
}\varphi \rangle \llangle\varphi \hspace{0.01in}|\hspace{0.01in}\mathcal{O}%
\rrangle  \notag \\
&=&\left[ 1+\llangle\mathcal{O}\hspace{0.01in}|\hspace{0.01in}\varphi %
\rrangle\right] ^{2}\langle \varphi \hspace{0.01in}|\hspace{0.01in}\varphi
\rangle +\llangle\mathcal{O}\hspace{0.01in}|\hspace{0.01in}\mathcal{O}%
\rrangle.\qquad  \label{composite}
\end{eqnarray}%
Here and below, a vertical bar is used to separate the (elementary or
composite) field of momentum $p$ (to the left) from the one of momentum $-p$
(to the right). The symbol $\llangle\cdots \rrangle$\ collects the
\textquotedblleft nonamputable\textquotedblright\ diagrams, which are those
that do not contain propagators of momentum $p$. The first equality of
identity (\ref{composite}) can be easily proved diagrammatically.

Formula (\ref{composite}) shows that the location of the pole is the same in 
$\langle \varphi \hspace{0.01in}|\hspace{0.01in}\varphi \rangle $ and $%
\langle \varphi ^{\prime }\hspace{0.01in}|\hspace{0.01in}\varphi ^{\prime
}\rangle $. We write it as $p^{2}=m_{\text{ph}}^{2}$, where $m_{\text{ph}}$
denotes the physical mass (possibly equipped with an imaginary part, if the
particle is unstable). On the other hand, the residue at the pole may
change. Precisely, we have%
\begin{equation*}
\langle \varphi \hspace{0.01in}|\hspace{0.01in}\varphi \rangle \simeq \frac{%
iZ}{p^{2}-m_{\text{ph}}^{2}+i\epsilon },\qquad \langle \varphi ^{\prime }%
\hspace{0.01in}|\hspace{0.01in}\varphi ^{\prime }\rangle \simeq \frac{%
iZ^{\prime }}{p^{2}-m_{\text{ph}}^{2}+i\epsilon },
\end{equation*}%
where $Z$ is the usual normalization factor and%
\begin{equation*}
Z^{\prime }=Z\left. \left[ 1+\llangle\mathcal{O}\hspace{0.01in}|\hspace{%
0.01in}\varphi \rrangle\right] \right\vert _{p^{2}=m_{\text{ph}}^{2}}^{2}
\end{equation*}
is the new normalization factor.

Now, consider the correlation functions that contain more than two $\varphi
^{\prime }$ insertions. Singling out one insertion at a time, the
diagrammatics easily gives 
\begin{equation*}
\langle \varphi ^{\prime }\hspace{0.03in}\text{rest}\rangle =\left[ 1+%
\llangle\mathcal{O}\hspace{0.01in}|\hspace{0.01in}\varphi \rrangle\right]
\langle \varphi \hspace{0.01in}|\hspace{0.01in}\varphi \rangle \llangle%
\varphi \hspace{0.03in}\text{rest}\rrangle+\llangle\mathcal{O}\hspace{0.03in}%
\text{rest}\rrangle,
\end{equation*}%
where the nonamputation $\llangle\cdots \rrangle$ only refers to the leg
under consideration. Thus, the identity%
\begin{equation}
\langle \prod\limits_{a=1}^{j}\frac{p_{a}^{2}-m_{\text{ph}}^{2}}{\sqrt{%
Z^{\prime }}}\varphi ^{\prime }(p_{a})\hspace{0.01in}\rangle _{\text{on-shell%
}}=\langle \hspace{0.01in}\prod\limits_{a=1}^{j}\frac{p_{a}^{2}-m_{\text{ph}%
}^{2}}{\sqrt{Z}}\varphi (p_{a})\rangle _{\text{on-shell}}  \label{onshellco}
\end{equation}%
holds, which proves that the $S$ matrix amplitudes do not change when we
make a (perturbative) change of field variables from $\varphi $ to $\varphi
^{\prime }$.

Applying this result to the cloud extension of Yang-Mills theory, we obtain,
in momentum space%
\begin{eqnarray}
&&\langle \prod\limits_{i=1}^{n}k_{i}^{2}\varepsilon _{i\text{d}}^{\mu
_{i}}(k_{i})A_{\mu _{i}\text{d}}(k_{i})\hspace{0.01in}\prod\limits_{a=1}^{j}%
\bar{u}_{s_{a}\text{d}}(p_{a})(\cancel{p}_{a}-\tilde{m})\psi _{\text{d}%
}(p_{a})\hspace{0.01in}\prod\limits_{b=1}^{j}\bar{\psi}_{\text{d}}(q_{b})(%
\slash\!\!\!{q}_{b}-\tilde{m})u_{s_{b}\text{d}}(q_{b})\rangle _{\text{%
on-shell}}  \notag \\
&&\quad =\langle \prod\limits_{i=1}^{n}k_{i}^{2}\varepsilon _{i}^{\mu
_{i}}(k_{i})A_{\mu _{i}}(k_{i})\hspace{0.01in}\prod\limits_{a=1}^{j}\bar{u}%
_{s_{a}}(p_{a})(\cancel{p}_{a}-\tilde{m})\psi (p_{a})\hspace{0.01in}%
\prod\limits_{b=1}^{j}\bar{\psi}(q_{b})(\slash\!\!\!{q}_{b}-\tilde{m}%
)u_{s_{b}}(q_{b})\rangle _{\text{on-shell}},\qquad  \label{onshecorr}
\end{eqnarray}%
where $\cancel{p}=\gamma ^{\mu }p_{\mu }$ and $\tilde{m}$ denotes the
physical mass. The polarizations $\varepsilon ^{\mu }(k)$ and $u_{s}(p)$
satisfy $k_{\mu }\varepsilon ^{\mu }(k)=0$ and $(\cancel{p}-\tilde{m}%
)u_{s}(p)=0$ and include the normalization factors $1/\sqrt{Z}$. The
\textquotedblleft dressed\textquotedblright\ polarizations $\varepsilon _{%
\text{d}}^{\mu }(k)$ and $u_{s\text{d}}(p)$ are the same, apart from having
normalization factors $1/\sqrt{Z^{\prime }}$. By the theorem proved in the
previous section, the right-hand side of (\ref{onshecorr}) is cloud
independent and coincides with the usual $S$ matrix amplitude.

Two differences between (\ref{onshellco}) and (\ref{onshecorr}) deserve to
be singled out. Formulas (\ref{dressedfields}) show that the expansion of $%
A_{\mu \text{d}}$ contains a linear contribution $-\partial _{\mu }\phi $,
besides $A_{\mu }$ itself, plus nonlinear terms. Thus, $A_{\mu \text{d}}$ is
not of the form $\varphi ^{\prime }=\varphi +\mathcal{O}(\varphi )$.
Nevertheless, the linear term $-\partial _{\mu }\phi $ becomes $ik_{\mu
}\phi (k)$, after the Fourier transform, and is killed by the polarization $%
\varepsilon ^{\mu }(k)$. This means that $\varepsilon _{\text{d}}^{\mu
}(k)A_{\mu \text{d}}(k)$ is of the required form, apart from an unimportant
normalization factor.

Second, we are not comparing correlation functions of the same theory, as in
(\ref{onshellco}). We are jumping from one theory (the extended one, to
which the left-hand side of (\ref{onshecorr}) refers) to another theory (the
non extended one, to which the right-hand side of (\ref{onshecorr}) refers).
This is possible, thanks to the result of the previous section.

In the end, the product $k^{2}\varepsilon _{\text{d}}^{\mu }(k)A_{\mu \text{d%
}}(k)$ is gauge invariant (and gauge independent, for the arguments we give
below) and its dressing is trivial: 
\begin{equation*}
\lim_{k^{2}\rightarrow 0}k^{2}\varepsilon _{\text{d}}^{\mu }(k)\langle
A_{\mu \text{d}}(k)\cdots \rangle =\lim_{k^{2}\rightarrow 0}k^{2}\varepsilon
^{\mu }(k)\langle A_{\mu }(k)\cdots \rangle .
\end{equation*}%
The same result holds for the fermions and any other elementary fields, if
present.

The identity (\ref{onshecorr}) proves that the ordinary theory of scattering
can be rephrased as a theory of scattering of dressed fields. We could even
forget about the ordinary fields altogether, and always work with dressed
fields, which have the advantage of being manifestly gauge invariant. In so
doing, both gauge invariance and gauge independence become manifest.

A straightforward consequence is that the usual $S$-matrix amplitudes are
gauge independent. With the usual methods, the proof of this result is
relatively simple in the Abelian case, but more demanding in the non-Abelian
one \cite{Kallosh}.

\section{Dressed correlation functions}

\label{cloudcorre}\setcounter{equation}{0}

In this section we study the correlation functions of the dressed fields. A
way to deal with their insertions systematically is by coupling them to new
sources and extending the generating functionals again. We replace the
action $S_{\text{tot}}$ inside (\ref{Z}) by%
\begin{equation}
S_{\text{tot}}^{\text{ext}}=S_{\text{tot}}+\int \left( J_{\text{d}}^{\mu
}A_{\mu \text{d}}+\bar{J}_{\psi \text{d}}\psi _{\text{d}}+\bar{\psi}_{\text{d%
}}J_{\psi \text{d}}\right) ,  \label{extra}
\end{equation}%
and denote the extended generating functionals by $Z_{\text{tot}}^{\text{ext}%
}(J,K,\tilde{J},\tilde{K},J_{\text{d}})=\exp (iW_{\text{tot}}^{\text{ext}%
}(J,K,$ $\tilde{J},\tilde{K},J_{\text{d}}))$. Note that the extended action
is gauge invariant, since (\ref{stot}) implies%
\begin{equation}
(S_{K}^{\text{gauge}},S_{\text{tot}}^{\text{ext}})=0.  \label{stotext}
\end{equation}

The correlation functions of the dressed fields are the functional
derivatives with respect to the dressed sources $J_{\text{d}}=(J_{\text{d}%
}^{\mu },\bar{J}_{\psi \text{d}},J_{\psi \text{d}})$.

\subsection{Gauge independence}

\label{gaugeproof}

It is straightforward to prove that the dressed correlation functions,
collected in the functional $Z_{\text{tot}}^{\text{ext}}(J_{\text{d}})=\exp
(iW_{\text{tot}}^{\text{ext}}(J_{\text{d}}))\equiv Z_{\text{tot}}^{\text{ext}%
}(0,0,0,0,J_{\text{d}})$, are gauge independent.

Assume that the gauge fermion $\Psi $ depends on some gauge-fixing parameter 
$\lambda $. A derivative with respect to $\lambda $ amounts to an insertion
of an $S_{K}^{\text{gauge}}$-exact functional:%
\begin{equation}
\frac{\partial }{\partial \lambda }W_{\text{tot}}^{\text{ext}}(J_{\text{d}%
})=\langle (S_{K}^{\text{gauge}},\Psi _{\lambda })\rangle _{J=K=\tilde{J}=%
\tilde{K}=0},  \label{gind1}
\end{equation}%
where $\Psi _{\lambda }=\partial \Psi /\partial \lambda $. We want to show
that the right-hand side of this identity vanishes.

Consider $\langle \Psi _{\lambda }\rangle _{J=K=\tilde{J}=\tilde{K}=0}$ and
perform a change of field variables 
\begin{equation*}
\delta \Phi ^{\alpha }=\theta (S_{K}^{\text{gauge}},\Phi ^{\alpha }),\qquad
\delta \tilde{\Phi}^{\alpha }=\theta (S_{K}^{\text{gauge}},\tilde{\Phi}%
^{\alpha }),
\end{equation*}%
in the functional integral that defines the numerator of the average.
Because of (\ref{stotext}), which also implies $(S_{K}^{\text{gauge}},\left.
S_{\text{tot}}^{\text{ext}}\right\vert _{K=\tilde{K}=0})=0$, everything is
invariant, but $\Psi _{\lambda }$. Thus,%
\begin{equation}
0=\langle \delta \Psi _{\lambda }\rangle _{J=K=\tilde{J}=\tilde{K}=0}=\theta
\langle (S_{K}^{\text{gauge}},\Psi _{\lambda })\rangle _{J=K=\tilde{J}=%
\tilde{K}=0},  \label{gind2}
\end{equation}%
as we wished to prove. Gauge independence will be verified explicitly in the
computations of the next sections. In section \ref{renormalization} we prove
that it survives the renormalization.

\section{Gauge/cloud duality}

\setcounter{equation}{0}\label{dual}

In this section we prove a \textit{gauge/cloud duality}, which relates the
gauge-trivial sector of the theory to the cloud sector.

We start by deriving the could transformation of the dressed gauge field
from (\ref{dLU}) and (\ref{dressedfields}). The result is 
\begin{equation}
(S_{K}^{\text{tot}},A_{\mu \text{d}})=(S_{K}^{\text{cloud}},A_{\mu \text{d}%
})=D_{\mu }(A_{\text{d}})H_{\text{d}},  \label{SAd}
\end{equation}%
where $D_{\mu }(A_{\text{d}})$ denotes the covariant derivative, evaluated
on the dressed field $A_{\text{d}}$. Instead, 
\begin{equation}
H_{\text{d}}\equiv -\frac{i}{g}(S_{K}^{\text{cloud}},\hat{U})\hat{U}^{-1}=%
\frac{i}{g}\hat{U}(S_{K}^{\text{cloud}},\hat{U}^{-1})=\frac{\mathrm{e}^{-ig%
\hspace{0.01in}\mathrm{ad}_{\phi }}-1}{-ig\hspace{0.01in}\mathrm{ad}_{\phi }}%
H  \label{Hd}
\end{equation}%
denotes the dressed cloud ghost $H$. In the last step of (\ref{Hd}) we have
used (\ref{dex}) with $X=ig\phi $. It is easy to check that $H_{\text{d}}$
is indeed gauge invariant, $(S_{K}^{\text{gauge}},H_{\text{d}})=0$. We see
that the could transformation of $A_{\mu \text{d}}$ is analogous to the
gauge transformation of $A_{\mu }$, provided the dressed fields replace the
undressed ones. By inverting (\ref{Hd}), we obtain%
\begin{equation}
H=R(-\phi ,H_{\text{d}}).  \label{HHd}
\end{equation}

Similarly, when we work out the cloud transformation of $H_{\text{d}}$, we
find that it mimics the gauge transformation of $C$:%
\begin{equation}
(S_{K}^{\text{cloud}},H_{\text{d}})=igH_{\text{d}}H_{\text{d}}.  \label{SHd}
\end{equation}

We can also introduce the dressed Faddeev-Popov ghosts%
\begin{equation}
C_{\text{d}}\equiv R(\phi ,C)=-\frac{ig\hspace{0.01in}\mathrm{ad}_{\phi }}{1-%
\mathrm{e}^{ig\hspace{0.01in}\mathrm{ad}_{\phi }}}C=(S_{K}^{\text{gauge}%
},\phi ),  \label{Cd}
\end{equation}%
which are clearly gauge invariant. Their cloud transformations read%
\begin{equation}
(S_{K}^{\text{cloud}},C_{\text{d}}^{a})=(S_{K}^{\text{gauge}},H^{a})=(S_{K}^{%
\text{gauge}},R^{a}(-\phi ,H_{\text{d}}))=-C_{\text{d}}^{b}\frac{\delta
R^{a}(\phi _{\text{d}},H_{\text{d}})}{\delta \phi _{\text{d}}^{b}},
\label{SKCd}
\end{equation}%
having defined $\phi _{\text{d}}=-\phi $ and used (\ref{HHd}).

Next, we consider the change of field variables%
\begin{equation}
\Phi ,\tilde{\Phi}\rightarrow \Phi _{\text{d}},\tilde{\Phi}_{\text{d}}
\label{redefa}
\end{equation}%
from undressed fields to dressed fields, by means of the definitions (\ref%
{dressedfields}), (\ref{Hd}), (\ref{Cd}) and $\phi _{\text{d}}=-\phi $,
leaving all the other fields unchanged: $\bar{C}_{\text{d}}=\bar{C}$, $B_{%
\text{d}}=B$, $\bar{H}_{\text{d}}=\bar{H}$ and $E_{\text{d}}=E$. The
transformations (\ref{redefa}) are perturbatively local, which means that
when we use them as changes of field variables in the functional integral,
the Jacobian determinant is equal to one (using the dimensional
regularization).

To ensure that all the properties derived so far continue to hold, we need
to preserve the antiparentheses. We can achieve this goal by embedding (\ref%
{redefa}) into a canonical transformation%
\begin{equation}
\Phi ,\tilde{\Phi},K,\tilde{K}\rightarrow \Phi _{\text{d}},\tilde{\Phi}_{%
\text{d}},K_{\text{d}},\tilde{K}_{\text{d}},  \label{canonical}
\end{equation}%
of the Batalin-Vilkovisky\ type. Its generating functional is%
\begin{equation*}
F(\Phi ,\tilde{\Phi},K_{\text{d}},\tilde{K}_{\text{d}})=\int \Phi _{\text{d}%
}(\Phi ,\tilde{\Phi})K_{\text{d}}+\int \tilde{\Phi}_{\text{d}}(\Phi ,\tilde{%
\Phi})\tilde{K}_{\text{d}}.
\end{equation*}

At the practical level, the whole operation amounts to work out the
transformations of the dressed fields, which we have already done, and
couple them to the dressed sources. Using (\ref{SAd}), (\ref{SHd}) and (\ref%
{SKCd}), we find 
\begin{eqnarray*}
S_{K}^{\text{gauge}} &=&\int C_{\text{d}}^{a}\tilde{K}_{\phi \text{d}%
}^{a}-\int B_{\text{d}}^{a}K_{\bar{C}\text{d}}^{a}, \\
S_{K}^{\text{cloud}} &=&-\int (D_{\mu }(A_{\text{d}})H_{\text{d}})^{a}K_{A%
\text{d}}^{\mu a}+\frac{g}{2}\int f^{abc}H_{\text{d}}^{b}H_{\text{d}}^{c}%
\tilde{K}_{H\text{d}}^{a}-i{g\int \bar{\psi}_{\text{d}}H_{\text{d}}K_{\bar{%
\psi}\text{d}}-ig\int K_{\psi \text{d}}H_{\text{d}}}\psi _{\text{d}} \\
&&-\int E_{\text{d}}^{a}\tilde{K}_{\bar{H}\text{d}}^{a}-\int R^{a}(\phi _{%
\text{d}},H_{\text{d}})\tilde{K}_{\phi \text{d}}^{a}+\int C_{\text{d}}^{b}%
\frac{\delta R^{a}(\phi _{\text{d}},H_{\text{d}})}{\delta \phi _{\text{d}%
}^{b}}\hspace{0.01in}K_{C\text{d}}^{a}.
\end{eqnarray*}%
We see that (\ref{canonical}) switches the gauge transformations and the
cloud transformations.

Similarly, it exchanges the roles of the gauge-fixing function $G^{a}$ and
the cloud function $V^{a}$: $G^{a}(A(\phi ,A_{\text{d}}))\leftrightarrow
V^{a}(A_{\text{d}})$. An important caveat of such an exchange is that it
understands that the prescription adopted for the gauge-trivial sector is
exchanged with the prescription adopted for the cloud sector.

For example, choosing the covariant gauge in (\ref{gferm}) and the covariant
cloud in (\ref{cloud fermion}), we have%
\begin{equation}
\Psi =\int \bar{C}_{\text{d}}^{a}\left( \partial ^{\mu }(\hat{U}_{\text{d}%
}A_{\mu \text{d}}\hat{U}_{\text{d}}^{-1}+\frac{i}{g}\hat{U}_{\text{d}%
}(\partial _{\mu }\hat{U}_{\text{d}}^{-1}))^{a}+\frac{\lambda }{2}B_{\text{d}%
}^{a}\right) ,\qquad \tilde{\Psi}=\int \bar{H}_{\text{d}}^{a}\left( \partial
^{\mu }A_{\mu \text{d}}^{a}+\frac{\tilde{\lambda}}{2}E_{\text{d}}^{a}\right)
,  \label{psid}
\end{equation}%
where $\hat{U}_{\text{d}}=\exp (-ig\phi _{\text{d}})=\hat{U}^{-1}$.

Collecting the various pieces together, the dual action reads%
\begin{equation}
S_{\text{tot}}=-\frac{1}{4}\int F_{\mu \nu }^{a}(A_{\text{d}})F^{\mu \nu
a}(A_{\text{d}})+\int \bar{\psi}_{\text{d}}(i\gamma ^{\mu }D_{\mu }(A_{\text{%
d}})-m)\psi _{\text{d}}+(S_{K}^{\text{tot}},\Psi +\tilde{\Psi})+S_{K}^{\text{%
tot}}.  \label{Sdual}
\end{equation}

Using the duality just proved, it is possible to simplify the calculations
of the correlation functions of the dressed fields. Actually, if we choose a
unique cloud for every insertion (see next section for the generalization to
multiclouds), the correlation functions of the dressed fields coincide with
the correlation functions of the undressed fields in a specific gauge.

This property can be proved by applying the canonical transformation (\ref%
{canonical}) to the dressed correlation functions. The result is an
identical correlation function where the dressed fields are replaced by the
undressed ones, the gauge-fixing is replaced by the cloud and the cloud is
replaced by the gauge-fixing. For example, if we use the covariant gauge (%
\ref{sgfconcr}) and the covariant cloud function (\ref{cloud}), we obtain 
\begin{eqnarray}
&&\langle A_{\mu _{1}\text{d}}(x_{1})\cdots \hspace{0.01in}A_{\mu _{n}\text{d%
}}(x_{n})\hspace{0.01in}\psi _{\text{d}}(y_{1})\cdots \psi _{\text{d}}(y_{j})%
\hspace{0.01in}\bar{\psi}_{\text{d}}(z_{1})\cdots \bar{\psi}_{\text{d}%
}(z_{j})\rangle  \notag \\
&&\qquad =\langle A_{\mu _{1}}(x_{1})\cdots \hspace{0.01in}A_{\mu
_{n}}(x_{n})\hspace{0.01in}\psi (y_{1})\cdots \psi (y_{j})\hspace{0.01in}%
\bar{\psi}(z_{1})\cdots \bar{\psi}(z_{j})\rangle _{\lambda \rightarrow 
\tilde{\lambda}}.  \label{lr}
\end{eqnarray}%
This property will be verified in the computations of the next sections. It
ensures that the left-hand side (which does not depend on $\lambda $ by
gauge independence), can be worked out by replacing $\lambda $ with $\tilde{%
\lambda}$ in the undressed correlation function appearing on the right-hand
side (which does not depend on $\tilde{\lambda}$ by cloud independence).

Typically, the left-hand side of (\ref{lr}) receives contributions from a
huge number of diagrams. However, the identity (\ref{lr}) implies that most
contributions cancel out in the end. For example, the two-point function of
the dressed gauge field amounts to just one diagram, if it is computed as
the right-hand side of (\ref{lr}), but tenths of diagrams if it is computed
as the left-hand side of (\ref{lr}).

\section{Multiclouds}

\setcounter{equation}{0}\label{multiclouds}

In this section we extend the formalism of the previous ones by adding
several copies of the could sector. This allows us to dress each insertion,
in a correlation function, with its own cloud, independently of the clouds
of the other insertions.

We introduce many cloud fields $\phi ^{i}$, where $i$ labels the copies,
together with their anticommuting partners $H^{i}$ (the cloud ghosts), the
antighosts $\bar{H}^{i}$ and the Lagrange multipliers $E^{i}$, collected in
the list $\tilde{\Phi}^{\alpha i}=(\phi ^{i},H^{i},\bar{H}^{i},E^{i})$. Then
we couple sources $\tilde{K}^{\alpha i}$ to their transformations, which
include the gauge transformations and the cloud transformations of each
copy. We collect them in the functionals 
\begin{eqnarray}
S_{K}^{\text{gauge}} &=&S_{K}-\sum_{i}\int R^{a}(\phi ^{i},C)\tilde{K}_{\phi
}^{ai}+\sum_{i}\int H^{bi}\frac{\delta R^{a}(\phi ^{i},C)}{\delta \phi ^{bi}}%
\hspace{0.01in}\tilde{K}_{H}^{ai},  \notag \\
S_{K}^{\text{cloud\hspace{0.01in}}i} &=&\int (H^{ai}\tilde{K}_{\phi
}^{ai}-E^{ai}\tilde{K}_{\bar{H}}^{ai}),\qquad S_{K}^{\text{cloud}%
}=\sum_{i}S_{K}^{\text{cloud\hspace{0.01in}}i},\qquad S_{K}^{\text{tot}%
}=S_{K}^{\text{gauge}}+S_{K}^{\text{cloud}}.\qquad  \label{Scloudi}
\end{eqnarray}%
Finally, we extend the definition (\ref{brackets}) of antiparentheses to
include all the copies:%
\begin{equation}
(X,Y)=\int \left[ \frac{\delta _{r}X}{\delta \Phi ^{\alpha }}\frac{\delta
_{l}Y}{\delta K^{\alpha }}-\frac{\delta _{r}X}{\delta K^{\alpha }}\frac{%
\delta _{l}Y}{\delta \Phi ^{\alpha }}+\sum_{i}\left( \frac{\delta _{r}X}{%
\delta \tilde{\Phi}^{\alpha i}}\frac{\delta _{l}Y}{\delta \tilde{K}^{\alpha
i}}-\frac{\delta _{r}X}{\delta \tilde{K}^{\alpha i}}\frac{\delta _{l}Y}{%
\delta \tilde{\Phi}^{\alpha i}}\right) \right] .  \label{BVi}
\end{equation}

It is easy to check that the identities (\ref{ss1}) and (\ref{ss2}) continue
to hold. The total cloud fermion can be just the sum of the cloud fermions
of each copy. We take\ 
\begin{equation}
\tilde{\Psi}(\Phi ,\tilde{\Phi})=\sum_{i}\tilde{\Psi}_{i},\qquad \tilde{\Psi}%
_{i}=\int \bar{H}^{ai}\left( V^{ai}+\frac{\tilde{\lambda}_{i}}{2}%
E^{ai}\right) ,  \label{psiti}
\end{equation}%
where $V^{i}$ are the gauge invariant cloud functions: $(S_{K}^{\text{gauge}%
},V^{i})=0$. For simplicity, we also assume that each $V^{i}$ depends on the 
$i$th cloud field $\phi ^{i}$ only (besides $A_{\mu }$), i.e., different
cloud sectors are not mixed by the cloud functions. It can be proved that
renormalization preserves the unmixing (see section \ref{renormalization}).

The total action of the extended theory is still (\ref{stota}),\ and
satisfies (\ref{cloudmaster}) and (\ref{stot}). Moreover,%
\begin{equation}
(S_{K}^{\text{cloud\hspace{0.01in}}i},S_{\text{tot}})=0  \label{SKtoti}
\end{equation}%
for every $i$.

We can always build gauge invariant functions with two cloud fields, since
the product $\hat{U}(\phi _{j})\hat{U}^{-1}(\phi _{k})\equiv \hat{U}_{jk}$
is gauge invariant for every $j$ and $k$. We have no powerful control on how
such functions propagate through the operations we make, once they are
turned on. The cloud unmixing just mentioned is an important simplification,
as long as we can prove that it is not ruined by renormalization and our own
manipulations.

The correlation functions that do not contain insertions of some cloud
sector are independent of that sector. Indeed, the proof of (\ref{ide}) can
be repeated for every sector separately. The multicloud propagators can be
easily derived using this property. Consider, for example, the case of two
clouds. Denote the cloud fields by $\phi _{1}$ and $\phi _{2}$ and choose
the cloud fermions (\ref{cloud fermion}) with parameters $\tilde{\lambda}%
_{1} $, $\tilde{\lambda}_{2}$, and the covariant cloud functions (\ref{cloud}%
). Finally, choose the covariant gauge (\ref{covg}). Then, the propagators (%
\ref{propa}) hold in each sector. In addition, we have%
\begin{equation}
\langle \phi ^{1}\hspace{0.01in}|\hspace{0.01in}\phi ^{2}\rangle =-\frac{1}{%
\tilde{\lambda}_{1}-\tilde{\lambda}_{2}}\left( \tilde{\lambda}_{2}\langle
\phi ^{1}\hspace{0.01in}|\hspace{0.01in}\phi ^{1}\rangle -\tilde{\lambda}%
_{1}\langle \phi ^{2}\hspace{0.01in}|\phi ^{2}\rangle \right) =-\frac{%
i\lambda \delta ^{ab}}{(p^{2}+i\epsilon )^{2}}.  \label{propa12}
\end{equation}%
The identity (\ref{propa12}) is easily proved from the sum%
\begin{equation*}
-\frac{1}{2\tilde{\lambda}_{1}}\int \left( \partial ^{\mu }A_{\text{d}\mu
}^{a1}\right) ^{2}-\frac{1}{2\tilde{\lambda}_{2}}\int \left( \partial ^{\mu
}A_{\text{d}\mu }^{a2}\right) ^{2},
\end{equation*}%
which shows that, at the quadratic level, the combination $\phi ^{1}-\phi
^{2}$ decouples from $A_{\mu }$ and from the combination $\tilde{\lambda}%
_{2}\phi ^{1}+\tilde{\lambda}_{1}\phi ^{2}$. This implies $\langle \phi
^{1}-\phi ^{2}\hspace{0.01in}|\hspace{0.01in}\tilde{\lambda}_{2}\phi ^{1}+%
\tilde{\lambda}_{1}\phi ^{2}\rangle _{0}=0$. Note that (\ref{propa12}) may
suggest that the cloud sectors mix. Nevertheless, renormalization does not
mix them, as shown in section \ref{renormalization}.

The correlation functions of dressed fields can be studied by means of the
extension%
\begin{equation}
S_{\text{tot}}^{\text{ext}}=S_{\text{tot}}+\sum_{i}\int \left( J_{\text{d}%
}^{\mu i}A_{\mu \text{d}}^{i}+\bar{J}_{\psi \text{d}}^{i}\psi _{\text{d}%
}^{i}+\bar{\psi}_{\text{d}}^{i}J_{\psi \text{d}}^{i}\right) ,  \label{sexti}
\end{equation}%
where $A_{\mu \text{d}}^{i}$, $\psi _{\text{d}}^{i}$ and $\bar{\psi}_{\text{d%
}}^{i}$ denote the dressed fields of the $i$th cloud sector.

The gauge/cloud duality is less powerful in the presence of many clouds. It
can be used to eliminate one cloud, or a combination of clouds, but not all
of them. For example, a correlation function 
\begin{equation}
\langle A_{\mu _{1}\text{d}}^{(1)}(x_{1})\cdots \hspace{0.01in}A_{\mu _{n}%
\text{d}}^{(n)}(x_{n})\hspace{0.01in}\psi _{\text{d}}^{(n+1)}(y_{1})\cdots
\psi _{\text{d}}^{(n+j)}(y_{j})\hspace{0.01in}\bar{\psi}_{\text{d}%
}^{(n+j+1)}(z_{1})\cdots \bar{\psi}_{\text{d}}^{(n+2j)}(z_{j})\rangle ,
\label{do}
\end{equation}%
with different clouds for every field, can be converted into%
\begin{equation}
\langle A_{\mu _{1}}(x_{1})A_{\mu _{2}\text{d}}^{(2)\hspace{0.01in}\prime
}(x_{1})\cdots \hspace{0.01in}A_{\mu _{n}\text{d}}^{(n)\hspace{0.01in}\prime
}(x_{n})\hspace{0.01in}\psi _{\text{d}}^{(n+1)\hspace{0.01in}\prime
}(y_{1})\cdots \psi _{\text{d}}^{(n+j)\hspace{0.01in}\prime }(y_{j})\hspace{%
0.01in}\bar{\psi}_{\text{d}}^{(n+j+1)\hspace{0.01in}\prime }(z_{1})\cdots 
\bar{\psi}_{\text{d}}^{(n+2j)\hspace{0.01in}\prime }(z_{j})\rangle ,
\label{reda}
\end{equation}%
by means of a canonical transformation of the form (\ref{canonical}), which
turns the first dressed field into its undressed version. The clouds of the
other insertions are redefined as a consequence. We have emphasized this by
means of primes in (\ref{reda}).

These operations preserve the unmixing, after further redefinitions of the
cloud fields themselves. Indeed, the transformation (\ref{canonical}) leads
to%
\begin{equation*}
A_{\mu _{i}\text{d}}^{(i)\hspace{0.01in}\prime }=\hat{U}_{i1}A_{\mu _{i}%
\text{d}}^{(1)}\hat{U}_{i1}^{-1}+\frac{i}{g}\hat{U}_{i1}(\partial _{\mu _{i}}%
\hat{U}_{i1}^{-1}),\qquad i>1.
\end{equation*}%
To restore the unmixing, after relabeling $A_{\mu \text{d}}^{(1)}$ as $%
A_{\mu }$, it is sufficient to define the new $i$th cloud field $\phi
_{i}^{\prime }(\phi _{i},\phi _{1})$, $i>1$, so as to have $\hat{U}%
_{i1}=\exp (-ig\phi _{i}^{\prime })$.

The proof that the usual correlation functions are cloud independent, given
in section \ref{cloudindep}, can be straightforwardly generalized to the
multicloud case. Similarly, the proof of formula (\ref{onshecorr}), which
states that the $S$ matrix amplitudes coincide with the usual ones, can be
generalized to the case where each insertion is dressed by means of its own,
independent cloud. Note that each insertion may require a different
normalization factor $1/\sqrt{Z^{\prime }}$, depending on the cloud.

\section{Absorptive parts}

\setcounter{equation}{0}\label{absorpt}

In this section we define the absorptive parts of the off-shell correlation
functions that contain insertions of dressed fields, and study their
properties.

If $S=1+iT$ denotes the $S$ matrix, the (amputated, connected) diagrams give 
$iT$ and the amplitudes are $T$. If the unitarity equation $SS^{\dag }=1$
holds, it implies $-2$Re$[iT]=TT^{\dag }\geqslant 0$. A virtue of the
identity $-2$Re$[iT]=TT^{\dag }$, known as optical theorem, is that it holds
diagram by diagram (which means: if we replace $iT$ by any diagram we want,
and $TT^{\dag }$ by a suitable sum of \textquotedblleft cut
diagrams\textquotedblright , built with \textquotedblleft cut
propagators\textquotedblright , the usual vertices and their complex
conjugates \cite{unitarity}). It also holds without putting the external
legs on shell. Moreover, as shown in ref. \cite{diagrammarMio}, it splits
into many independent, purely algebraic spectral optical identities, because
different thresholds do not talk to one another.

It also holds with non amputated diagrams. To see this, it is sufficient to
attach fictitious vertices to the legs that we do not want to amputate. At
the practical level, this amounts to multiplying each of them by a factor $%
-i $. The identity $-2$Re$[iT]=TT^{\dag }$ also holds with insertions of
local composite fields (which can be attached to other fictitious vertices
-- for this reason, each of them brings a further factor $-i$). Combining
elementary and composite fields, the identity also holds with insertions of
dressed fields (each of which must be multiplied by $-i$).

We define the absorptive part of an off-shell correlation function, with or
without insertions of dressed fields, as minus twice the real part of its
amputated version, multiplied by the polarizations and the normalization
factors $\sqrt{Z}$ of the external states. It is expected to be nonnegative
if the unitarity equation $SS^{\dag }=1$ holds, by the arguments given
above. In our calculations, the factors $\sqrt{Z}$ can be set to one, since
the absorptive parts we are going to compute vanish at the tree level. Note
that more polarizations may be allowed off shell than on shell.

The extended action $S_{\text{tot}}^{\text{ext}}$ is local and Hermitian.
However, formula (\ref{cloudkin}) shows that the cloud fields $\phi ^{i}$
(which are dimensionless) do not have ordinary kinetic terms, but
higher-derivative ones. For this reason, unitarity and the diagrammatic
optical theorem are guaranteed to hold only if we use the fakeon
prescription and projection for the cloud fields (see below). If not, we
expect to find unphysical absorptive parts. The results of our computations
confirm these claims.

\section{Dressed fermion self-energy}

\setcounter{equation}{0}\label{drfermion}

In this section and the next two we illustrate the properties proved so far
in explicit calculations. We concentrate on the two-point functions of the
gauge fields and the fermions to order $g^{2}$, which means one loop. For
simplicity, we use the same cloud for all the insertions. We compare several
types of clouds, gauge-fixings and prescriptions.

The four-leg vertices, which are multiplied by $g^{2}$, contribute only to
tadpoles, which vanish using the dimensional regularization. Ignoring them,
it is sufficient to expand the dressed fields (\ref{dressedfields}) and the
cloud action to order $g$. We find%
\begin{equation*}
A_{\mu \text{d}}=A_{\mu }-\partial _{\mu }\phi -\frac{ig}{2}[\phi ,2A_{\mu
}-\partial _{\mu }\phi ]+\mathcal{O}(g^{2}),\qquad \psi _{\text{d}}=\psi
-ig\phi \psi +\mathcal{O}(g^{2}).
\end{equation*}%
For the moment, we concentrate on the covariant clouds and the covariant
gauge-fixings, and use the Feynman $i\epsilon $ prescription everywhere. The
cloud action (\ref{sKpsit}) reads%
\begin{eqnarray}
(S_{K}^{\text{cloud}},\tilde{\Psi}) &&=-\frac{1}{2\tilde{\lambda}}\int
\left( \partial ^{\mu }A_{\mu }^{a}-\square \phi ^{a}\right) ^{2}+\frac{g}{2%
\tilde{\lambda}}f^{abc}\int \left( \partial ^{\mu }\partial ^{\nu }A_{\mu
}^{a}-\square \partial ^{\nu }\phi ^{a}\right) \phi ^{b}(2A_{\nu
}^{c}-\partial _{\nu }\phi ^{c})  \notag \\
&&\!\!\!\!\!\!\!\!\!\!\!\!\!\!\!\!\!\!\!\!\!\!{\ +\int (\partial ^{\mu }\bar{%
H}^{a})(\partial _{\mu }H^{a})-\frac{g}{2}f^{abc}\int \left[ (\square \bar{H}%
^{a})\phi ^{b}-2(\partial ^{\mu }\bar{H}^{a})(A_{\mu }^{b}-\partial _{\mu
}\phi ^{b})\right] H^{c}+\mathcal{O}(g^{2}).}  \label{scloud2}
\end{eqnarray}

\begin{figure}[t]
\begin{center}
\includegraphics[width=12truecm]{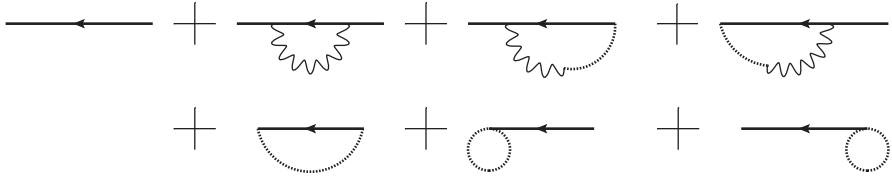}
\end{center}
\caption{Two-point function of the dressed fermion to order $g^{2}$. The
dashed line denotes the cloud field $\protect\phi $}
\label{dressedfermion}
\end{figure}

The two-point function of the dressed gauge fields is calculated in section %
\ref{gauge2}. Here we concentrate on the two-point function of the dressed
fermion, which reads 
\begin{equation}
\langle \psi _{\text{d}}\hspace{0.01in}|\hspace{0.01in}\bar{\psi}_{\text{d}%
}\rangle =\langle \psi \hspace{0.01in}|\hspace{0.01in}\bar{\psi}\rangle
-ig\langle \phi \psi \hspace{0.01in}|\hspace{0.01in}\bar{\psi}\rangle
+ig\langle \psi \hspace{0.01in}|\hspace{0.01in}\bar{\psi}\phi \rangle
+g^{2}\langle \phi \psi \hspace{0.01in}|\hspace{0.01in}\bar{\psi}\phi
\rangle -\frac{g^{2}}{2}\langle \phi ^{2}\psi \hspace{0.01in}|\hspace{0.01in}%
\bar{\psi}\rangle -\frac{g^{2}}{2}\langle \psi \hspace{0.01in}|\hspace{0.01in%
}\bar{\psi}\phi ^{2}\rangle +\mathcal{O}(g^{3}).  \label{fdfd}
\end{equation}%
The diagrams contributing to the right-hand side of (\ref{fdfd}) are shown
in fig. \ref{dressedfermion}. 
We restrict to the massless limit, which makes the formulas more explicit.

The last two diagrams\ of fig. \ref{dressedfermion} are tadpoles, which
vanish using the dimensional regularization. We have included them just to
show that $\langle \psi _{\text{d}}\hspace{0.01in}|\hspace{0.01in}\bar{\psi}%
_{\text{d}}\rangle $ in not plagued by infrared problems, no matter what
regularization we use. The reason is that the denominators $\sim
1/(p^{2})^{2}$ (which give potential infrared divergences with a generic
regularization technique) cancel out among the last three diagrams. Their
cancellation can be verified by taking the loop momentum to zero. In that
limit, a fermion propagator $i\gamma ^{\mu }p_{\mu }/(p^{2}+i\epsilon )$
factors out and all the diagrams become identical tadpoles. Since last two
are multiplied by $-g^{2}/2$, while the third to last one is multiplied by $%
g^{2}$, the total vanishes. We have not included other tadpole diagrams,
because they vanish identically (they factorize the trace of $T^{a}$ or a
contraction like $f^{abb}$).

The second diagram is the ordinary fermion self-energy. Added to the tree
propagator, it gives%
\begin{equation}
\langle \psi \hspace{0.01in}|\hspace{0.01in}\bar{\psi}\rangle =\frac{i\gamma
^{\mu }p_{\mu }}{p^{2}+i\epsilon }\left[ 1-\frac{g^{2}\lambda }{8\pi
^{2}\varepsilon }\frac{N_{c}^{2}-1}{2N_{c}}(-p^{2}-i\epsilon )^{-\varepsilon
/2}\right] +\mathcal{O}(g^{4}),  \label{psipsi}
\end{equation}%
where $\varepsilon =4-D$ and $D$ is the continued spacetime dimension. The
remaining diagrams of fig. \ref{dressedfermion} give%
\begin{equation}
\frac{i\gamma ^{\mu }p_{\mu }}{p^{2}+i\epsilon }\frac{g^{2}(\lambda -\tilde{%
\lambda})}{8\pi ^{2}\varepsilon }\frac{N_{c}^{2}-1}{2N_{c}}(-p^{2}-i\epsilon
)^{-\varepsilon /2}+\mathcal{O}(g^{4}),  \label{psipsi2}
\end{equation}%
so in total we get%
\begin{equation}
\langle \psi _{\text{d}}\hspace{0.01in}|\hspace{0.01in}\bar{\psi}_{\text{d}%
}\rangle =\frac{i\gamma ^{\mu }p_{\mu }}{p^{2}+i\epsilon }\left[ 1-\frac{%
g^{2}\tilde{\lambda}}{8\pi ^{2}\varepsilon }\frac{N_{c}^{2}-1}{2N_{c}}%
(-p^{2}-i\epsilon )^{-\varepsilon /2}\right] +\mathcal{O}(g^{4}).
\label{psidpsid}
\end{equation}%
The dependence on the gauge-fixing parameter $\lambda $ has disappeared, as
expected. Nevertheless, the result depends on the choice of the cloud,
through the parameter $\tilde{\lambda}$.

We see that 
\begin{equation*}
\langle \psi _{\text{d}}\hspace{0.01in}|\hspace{0.01in}\bar{\psi}_{\text{d}%
}\rangle =\langle \psi \hspace{0.01in}|\hspace{0.01in}\bar{\psi}\rangle
_{\lambda \rightarrow \tilde{\lambda}},
\end{equation*}%
in agreement with (\ref{lr}).

The renormalization of $\langle \psi _{\text{d}}\hspace{0.01in}|\hspace{%
0.01in}\bar{\psi}_{\text{d}}\rangle $ requires the counterterm 
\begin{equation}
\frac{g^{2}\tilde{\lambda}}{8\pi ^{2}\varepsilon }\frac{N_{c}^{2}-1}{2N_{c}}%
\int \bar{J}_{\psi \text{d}}\frac{i\gamma ^{\mu }\partial _{\mu }}{\square
-i\epsilon }J_{\psi \text{d}}.  \label{noc}
\end{equation}%
However, there is no need to insert it explicitly into the action $S_{\text{%
tot}}^{\text{ext}}$ of (\ref{extra}). As the nonlocal nature of (\ref{noc})
suggests, (\ref{noc}) is generated automatically by the Legendre transform
that relates the generating functional $\Gamma _{\text{tot}}^{\text{ext}}$
of the one-particle irreducible diagrams to the generating functional $W_{%
\text{tot}}^{\text{ext}}$ of the connected Green functions.

After renormalization, we take $\varepsilon $ to zero and find%
\begin{equation}
\langle \psi _{\text{d}}\hspace{0.01in}|\hspace{0.01in}\bar{\psi}_{\text{d}%
}\rangle _{R}=\frac{i\gamma ^{\mu }p_{\mu }}{p^{2}+i\epsilon }\left[ 1+\frac{%
g^{2}\tilde{\lambda}}{(4\pi )^{2}}\frac{N_{c}^{2}-1}{2N_{c}}\ln
(-p^{2}-i\epsilon )\right] +\mathcal{O}(g^{4}).  \label{psudR}
\end{equation}

Now we extract the absorptive part of this expression. Since we have
prescribed every field \`{a} la Feynman, the arguments of the previous
section alert us that the result might be unphysical.

Recall that we are working in the massless limit. We study the sign of the
absorptive part by summing on the external polarization states $u_{s}$. We
use the identity $\sum_{s}u_{s}\bar{u}_{s}=\gamma ^{\mu }p_{\mu }$ without
assuming $p^{2}=0$, because we stay off shell. This amounts to choosing the
states $u_{s}=\sqrt{p^{0}}(0,\ldots 1,\ldots ,0)^{t}$ (with 1 in position $s$
and zeros elsewhere) and the basis $\gamma ^{\mu }=((0,\sigma ^{\mu }),(\bar{%
\sigma}^{\mu },0))^{t}$, $\sigma ^{\mu }=(1,\mathbf{\sigma })$, $\bar{\sigma}%
^{\mu }=(1,-\mathbf{\sigma })$, where $^{t}$ denotes the transpose, $p^{0}>0$
and $s=1,2,3,4$. Then, the absorptive part of $\langle \psi _{\text{d}}%
\hspace{0.01in}|\hspace{0.01in}\bar{\psi}_{\text{d}}\rangle _{R}$ is%
\begin{equation}
\text{Abso}[\langle \psi _{\text{d}}\hspace{0.01in}|\hspace{0.01in}\bar{\psi}%
_{\text{d}}\rangle _{R}]=-2\text{Re}\sum_{s}\bar{u}_{s}(-i\gamma ^{\mu
}p_{\mu })\langle \psi _{\text{d}}\hspace{0.01in}|\hspace{0.01in}\bar{\psi}_{%
\text{d}}\rangle _{R}(-i\gamma ^{\nu }p_{\nu })u_{s}=\frac{g^{2}\tilde{%
\lambda}}{2\pi }\frac{N_{c}^{2}-1}{2N_{c}}p^{2}\theta (p^{2})+\mathcal{O}%
(g^{4}).  \label{absopsi}
\end{equation}%
A factor 4 comes from the spinor trace.

We see that the sign of (\ref{absopsi}) is positive or negative, depending
on the sign of the cloud parameter $\tilde{\lambda}$. As already remarked,
the could field $\phi $ has higher-derivative kinetic terms. Thus, if we
quantize it by means of the Feynman prescription, as we have done so far, it
can propagate ghosts and violate the optical theorem.

We need to pay more attention to the clouds we choose, otherwise they can
inject unphysical degrees of freedom into the theory and make the
computations of the dressed correlation functions uninteresting.

\section{Purely virtual clouds}

\setcounter{equation}{0}\label{pvc}

The clouds we have been using so far are physically unacceptable, because
they add degrees of freedom that do not belong to the fundamental theory. A
way to preserve the content of the fundamental theory is to switch to purely
virtual clouds. This way, we can use the correlation functions of the
dressed fields as tools to extract the physical content of the off-shell
correlation functions of the fundamental fields.

For a clearer understanding of what is going on, it may be helpful to
introduce the special gauge of ref. \cite{Unitaritycc}, which is defined by
the gauge function $G^{a}(A)=\lambda \partial _{0}A^{0a}+\mathbf{\nabla }%
\cdot \mathbf{A}^{a}$ in (\ref{gferm}), where $A^{\mu a}=(A^{0a},\mathbf{A}%
^{a})$. The gauge fermion reads%
\begin{equation}
\Psi (\Phi )=\int \bar{C}^{a}\left( \lambda \partial _{0}A^{0a}+\mathbf{%
\nabla }\cdot \mathbf{A}^{a}+\frac{\lambda }{2}B^{a}\right) .  \label{spepsi}
\end{equation}

For a more direct gauge/cloud duality and a convenient switch back and forth
between the gauge-trivial sector and the cloud sector, we mimic the special
gauge into a \textquotedblleft special cloud\textquotedblright , by choosing
the cloud function $V^{a}(A,\phi )=\tilde{\lambda}\partial _{0}A_{\text{d}%
}^{0a}+\mathbf{\nabla }\cdot \mathbf{A}_{\text{d}}^{a}$, instead of (\ref%
{cloud}). Then the cloud fermion (\ref{cloud fermion}) reads 
\begin{equation}
\tilde{\Psi}(\Phi ,\tilde{\Phi})=\int \bar{H}^{a}\left( \tilde{\lambda}%
\partial _{0}A_{\text{d}}^{0a}+\mathbf{\nabla }\cdot \mathbf{A}_{\text{d}%
}^{a}+\frac{\tilde{\lambda}}{2}E^{a}\right) .  \label{spepsit}
\end{equation}

With these choices the propagators of the gauge fields and the cloud field
become, after integrating the Lagrange multipliers $B$ and $E$ out,%
\begin{eqnarray}
\langle A^{ia}(p)\hspace{0.01in}A^{jb}(-p)\rangle _{0} &=&\frac{i\delta
^{ab}\Pi ^{ij}}{p^{2}+i\epsilon }+\frac{i\lambda \delta ^{ab}p^{i}p^{j}}{%
\mathbf{p}^{2}(\hat{p}\cdot p)},\qquad \langle A^{0a}(p)\hspace{0.01in}%
A^{0b}(-p)\rangle _{0}=-\frac{i\delta ^{ab}}{\hat{p}\cdot p},  \notag \\
\langle A^{\mu a}(p)\hspace{0.01in}\phi ^{b}(-p)\rangle _{0} &=&-\frac{%
\delta ^{ab}(\tilde{\lambda}E,\lambda \mathbf{p})}{(\tilde{p}\cdot p)(\hat{p}%
\cdot p)},\quad \langle \phi ^{a}(p)\hspace{0.01in}\phi ^{b}(-p)\rangle
_{0}=-\frac{i\delta ^{ab}(\lambda +\tilde{\lambda})}{(\tilde{p}\cdot p)(\hat{%
p}\cdot p)},\qquad  \label{propaspecial}
\end{eqnarray}%
together with $\langle A^{ia}(p)\hspace{0.01in}A^{0b}(-p)\rangle _{0}=0$,
where $\Pi ^{ij}=\delta ^{ij}-(p^{i}p^{j}/\mathbf{p}^{2})$, $p^{\mu
}=(p^{0},p^{i})=(E,\mathbf{p})$, $\hat{p}^{\mu }=(\lambda E,\mathbf{p})$ and 
$\tilde{p}^{\mu }=(\tilde{\lambda}E,\mathbf{p})$. The ghost propagators are%
\begin{equation*}
\langle C^{a}(p)\hspace{0.01in}\bar{C}^{b}(-p)\rangle _{0}=\frac{i\delta
^{ab}}{\hat{p}\cdot p},\qquad \langle H^{a}(p)\hspace{0.01in}\bar{H}%
^{b}(-p)\rangle _{0}=\frac{i\delta ^{ab}}{\tilde{p}\cdot p}.
\end{equation*}

We have left the denominators $1/(\hat{p}\cdot p)$ and $1/(\tilde{p}\cdot p)$
unprescribed. The former belong to the gauge-trivial sector, while the
latter belong to the cloud sector. Obviously, the poles $1/p^{2}$ belong to
the physical sector. The virtue of the special gauge, combined with the
special cloud, is that it keeps the three sectors distinct throughout the
calculations (at generic $\lambda $ and $\tilde{\lambda}$). The distinction
also holds through the threshold decomposition of \cite{diagrammarMio},
which is crucial to define the diagrammatics of purely virtual particles.
Specifically, the physical thresholds, which are those originated solely by
the physical poles $1/p^{2}$, are kept distinct from the unphysical
thresholds, which are those that receive any contributions from the poles $%
1/(\hat{p}\cdot p)$ and $1/(\tilde{p}\cdot p)$. Also note that at generic $%
\lambda $ and $\tilde{\lambda}$ there are no double poles (which is what
makes the special gauge \textquotedblleft special\textquotedblright\ \cite%
{Unitaritycc}).

Since the physical quantities are gauge independent, it does not matter
which prescription (e.g., Feynman $i\epsilon $, or purely virtual) we use
for the poles $1/(\hat{p}\cdot p)$ of the gauge-trivial sector, as long as
they are all prescribed the same way. The poles $1/(\tilde{p}\cdot p)$
belonging to the cloud sector, instead, should be quantized as purely
virtual, according to the rules of \cite{diagrammarMio}.

Practically, this means that we start with the Feynman prescription
everywhere, as we would normally do, then make the threshold decomposition
of ref. \cite{diagrammarMio}, and finally drop every $\tilde{\lambda}$%
-dependent threshold. These operations render the whole cloud sector purely
virtual.

For a variety of applications (and to have a more direct gauge/cloud match),
it may be convenient to work with a purely virtual gauge-trivial sector as
well. To do so, it is sufficient to adopt the fakeon prescription for the
poles $1/(\hat{p}\cdot p)$ as well. This option was introduced in \cite%
{PVGauge} to provide a more direct proof of unitarity in gauge theories.

At the end, we just drop all the $\lambda $, $\tilde{\lambda}$-dependent
thresholds that we find in the decomposition of ref. \cite{diagrammarMio},
and keep only the physical ones. What we obtain is a powerful physical gauge.

\subsection{Dressed fermion self-energy, again}

We use the framework just defined to calculate the dressed fermion
self-energy anew. For simplicity, we calculate it at rest, and assume $%
\lambda >0$, $\tilde{\lambda}>0$. First, we use the Feynman prescription for
every pole. Then, we describe what changes when we use purely virtual clouds.

The undressed two-point function turns out to be%
\begin{equation}
\langle \psi \hspace{0.01in}|\hspace{0.01in}\bar{\psi}\rangle =\frac{i\gamma
^{\mu }p_{\mu }}{p^{2}+i\epsilon }\left[ 1-\frac{g^{2}(1-\sqrt{\lambda }%
+2\lambda )}{8\pi ^{2}\varepsilon (1+\sqrt{\lambda })}\frac{N_{c}^{2}-1}{%
2N_{c}}(-p^{2}-i\epsilon )^{-\varepsilon /2}\right] +\mathcal{O}(g^{4}),
\label{imaF}
\end{equation}%
instead of (\ref{psipsi}). The other diagrams of fig. \ref{dressedfermion}
give%
\begin{equation*}
-\frac{i\gamma ^{\mu }p_{\mu }}{p^{2}+i\epsilon }\frac{g^{2}(\sqrt{\lambda }-%
\sqrt{\tilde{\lambda}})(1-\sqrt{\lambda }-\sqrt{\tilde{\lambda}}-\sqrt{%
\lambda \tilde{\lambda}})}{4\pi ^{2}\varepsilon (1+\sqrt{\lambda })(1+\sqrt{%
\tilde{\lambda}})}\frac{N_{c}^{2}-1}{2N_{c}}(-p^{2}-i\epsilon
)^{-\varepsilon /2}+\mathcal{O}(g^{4}),
\end{equation*}%
instead of (\ref{psipsi2}). In total we get 
\begin{equation*}
\langle \psi _{\text{d}}\hspace{0.01in}|\hspace{0.01in}\bar{\psi}_{\text{d}%
}\rangle =\frac{i\gamma ^{\mu }p_{\mu }}{p^{2}+i\epsilon }\left[ 1-\frac{%
g^{2}(1-\sqrt{\tilde{\lambda}}+2\tilde{\lambda})}{8\pi ^{2}\varepsilon (1+%
\sqrt{\tilde{\lambda}})}\frac{N_{c}^{2}-1}{2N_{c}}(-p^{2}-i\epsilon
)^{-\varepsilon /2}\right] +\mathcal{O}(g^{4})=\langle \psi \hspace{0.01in}|%
\hspace{0.01in}\bar{\psi}\rangle _{\lambda \rightarrow \tilde{\lambda}},
\end{equation*}%
in agreement with formula (\ref{lr}). Note that although formula (\ref{lr})
was derived in the covariant gauge and with a covariant cloud, it also
applies to the present calculation, because the exchange $%
G^{a}(A)\leftrightarrow V^{a}(\phi ,A)$ still amounts to $\lambda
\leftrightarrow \tilde{\lambda}$, for the choices of gauge-fixing and cloud
that we have made.

Now we switch to purely virtual clouds, using the fakeon prescription for
the poles $1/(\tilde{p}\cdot p)$ of (\ref{propaspecial}). The imaginary part
of $\langle \psi _{\text{d}}\hspace{0.01in}|\hspace{0.01in}\bar{\psi}_{\text{%
d}}\rangle $ remains the same, so we can focus on the absorptive part%
\begin{equation}
\text{Abso}[\langle \psi _{\text{d}}\hspace{0.01in}|\hspace{0.01in}\bar{\psi}%
_{\text{d}}\rangle _{R}]=-2\text{Re}\sum_{s}\bar{u}_{s}(-i\gamma ^{\mu
}p_{\mu })\langle \psi _{\text{d}}\hspace{0.01in}|\hspace{0.01in}\bar{\psi}_{%
\text{d}}\rangle _{R}(-i\gamma ^{\nu }p_{\nu })u_{s}=2p^{2}\text{Re}\text{Tr}%
[\gamma ^{\mu }p_{\mu }\langle \psi _{\text{d}}\hspace{0.01in}|\hspace{0.01in%
}\bar{\psi}_{\text{d}}\rangle _{R}],  \label{abso}
\end{equation}%
which is the one affected by the prescription. By definition, all the
contributions to (\ref{abso}) coming from the $\tilde{\lambda}$-dependent
thresholds drop out, due to the fakeon prescription for the poles $1/(\tilde{%
p}\cdot p)$. The contributions of the $\lambda $-dependent thresholds
compensate one another, so we can use the prescription we want for the poles 
$1/(\hat{p}\cdot p)$. Choosing the fakeon prescription for them as well, we
see that the absorptive part of $\langle \psi _{\text{d}}\hspace{0.01in}|%
\hspace{0.01in}\bar{\psi}_{\text{d}}\rangle $ comes from the sole second
diagram of fig. \ref{dressedfermion}, which is the usual self-energy
diagram, provided we restrict the gauge-field propagator (\ref{propaspecial}%
) to its physical part 
\begin{equation}
\langle A^{\mu a}(p)\hspace{0.01in}A^{\nu b}(-p)\rangle _{0\text{phys}}=i%
\frac{\delta ^{ab}\delta _{i}^{\mu }\delta _{j}^{\nu }\Pi ^{ij}}{%
p^{2}+i\epsilon },  \label{AAphys}
\end{equation}%
which is the one obtained by dropping the poles $1/(\hat{p}\cdot p)$ and $1/(%
\tilde{p}\cdot p)$ in (\ref{propaspecial}). Finally, the absorptive part
does not need renormalization. The final result is%
\begin{equation}
\text{Abso}[\langle \psi _{\text{d}}\hspace{0.01in}|\hspace{0.01in}\bar{\psi}%
_{\text{d}}\rangle _{R}]=\frac{g^{2}}{2\pi }p^{2}\theta (p^{2}).
\label{absopsid}
\end{equation}%
As desired, it is gauge independent, cloud independent and positive.
Ultimately, this is the physical content of the fermion two-point function
at one loop.

The calculation has been done for fermions at rest. The general, off-shell
result is not Lorentz invariant. The reason is that, in order to compensate
for the gauge dependence of the undressed fermion $\psi $, the cloud must be
built with the longitudinal and temporal components of the gauge fields. The
very definition of such components requires to specify a Lorentz frame. If
we want, we can even choose different Lorentz frames for each cloud and for
the gauge-fixing.

The cloud Faddeev-Popov determinant did not contribute so far. It
contributes to $\langle \psi _{\text{d}}\hspace{0.01in}|\hspace{0.01in}\bar{%
\psi}_{\text{d}}\rangle $ starting from two loops. It also contributes to
the one-loop two-point function $\langle A_{\text{d}}\hspace{0.01in}|\hspace{%
0.01in}A_{\text{d}}\rangle $ of the dressed gauge fields (see below).
Clearly, it is crucial for the gauge/cloud duality.

Sometimes, it may be convenient to simplify the calculations by choosing $%
\lambda =\tilde{\lambda}=1$. In that case, the propagators (\ref%
{propaspecial}) become 
\begin{eqnarray}
\langle A^{ia}(p)\hspace{0.01in}A^{jb}(-p)\rangle _{0} &=&\frac{i\delta
^{ab}\Pi ^{ij}}{p^{2}+i\epsilon }+\left. \frac{i\delta ^{ab}p^{i}p^{j}}{%
\mathbf{p}^{2}p^{2}}\right\vert _{\text{f}},\qquad \langle A^{0a}(p)\hspace{%
0.01in}A^{0b}(-p)\rangle _{0}=-\left. \frac{i\delta ^{ab}}{p^{2}}\right\vert
_{\text{f}},  \notag \\
\langle A^{\mu a}(p)\hspace{0.01in}\phi ^{b}(-p)\rangle _{0} &=&-\left. 
\frac{\delta ^{ab}p^{\mu }}{(p^{2})^{2}}\right\vert _{\text{f}},\quad
\langle \phi ^{a}(p)\hspace{0.01in}\phi ^{b}(-p)\rangle _{0}=-\left. \frac{%
2i\delta ^{ab}}{(p^{2})^{2}}\right\vert _{\text{f}},  \label{PVspecial}
\end{eqnarray}%
where the subscript \textquotedblleft f\textquotedblright\ denotes the
fakeon prescription. The various types of thresholds (physical, gauge or
cloud) are not manifestly distinct at $\lambda =\tilde{\lambda}=1$, so we
have to keep track of their origins in different ways.

Note that (\ref{PVspecial}) involves square denominators like $1/(p^{2})^{2}$%
. Their fakeon prescription in a diagram $G$ is defined as follows: 
\begin{equation*}
G\left[ \left. \frac{1}{(p^{2})^{2}}\right\vert _{\text{f}}\right]
\rightarrow -\lim_{\mu ^{2}\rightarrow 0}\frac{\mathrm{d}}{\mathrm{d}\mu ^{2}%
}G\left[ \left. \frac{1}{p^{2}-\mu ^{2}}\right\vert _{\text{f}}\right] .
\end{equation*}

Power counting is straightforward with both the covariant and special gauge
and clouds. It may not work equally well with other choices of gauges and
clouds. An example is the Coulomb gauge, which can be obtained from the
special gauge by letting $\lambda $ tend to zero. Similarly, the
\textquotedblleft Coulomb cloud\textquotedblright\ can be obtained by
letting $\tilde{\lambda}$ tend to zero in the special cloud. In those
limits, the integrals on the loop energies and the integrals on the space
components of the loop momenta obey different power counting rules.

No particular prescription in needed to treat the Coulomb poles $1/\mathbf{p}%
^{2}$. They are purely virtual by accident, in some sense. Furthermore,
formula (\ref{lr}) shows that the correlation functions of the dressed
fields at $\tilde{\lambda}=0$ coincide with those of the undressed fields in
the Coulomb gauge. The absorptive part at rest clearly coincides with (\ref%
{absopsid}), because it is cloud independent.

\section{Dressed gauge-field two-point function}

\setcounter{equation}{0}\label{gauge2}

In this section we study the dressed gauge-field two-point function at one
loop, sticking to pure Yang-Mills theory for simplicity. As before, we start
from the covariant gauge and the covariant cloud, with the Feynman
prescription everywhere. At a second stage we switch to purely virtual
clouds.

First, we verify that the ordinary two-point function $\langle A_{\mu }^{a}%
\hspace{0.01in}|\hspace{0.01in}A_{\nu }^{b}\rangle $ is cloud independent,
to check the results of section \ref{cloudindep}. The diagrams contributing
to $\langle A_{\mu }^{a}\hspace{0.01in}|\hspace{0.01in}A_{\nu }^{b}\rangle $
are too many to be listed here and include loops of cloud ghosts $H$-$\bar{H}
$. Collecting everything together and subtracting the divergent part, the
result is the same as usual, i.e.,%
\begin{equation}
\langle A^{\mu a}\hspace{0.01in}|\hspace{0.01in}A^{\nu b}\rangle _{\text{%
one-loop}}=\frac{ig^{2}N_{c}(13-3\lambda )}{6(4\pi )^{2}(p^{2})^{2}}\delta
^{ab}(p^{2}\eta ^{\mu \nu }-p^{\mu }p^{\nu })\ln (-p^{2}-i\epsilon ).
\label{r1}
\end{equation}%
As expected, the dependence on the cloud parameter $\tilde{\lambda}$
disappears and the gauge dependence remains.

Next, we compute the two-point function $\langle A_{\text{d}\mu }^{a}\hspace{%
0.01in}|\hspace{0.01in}A_{\text{d}\nu }^{b}\rangle $ of the dressed gauge
fields, still in the covariant gauge. The number of diagrams is even larger,
but the final result is extremely simple and coincides with (\ref{r1}),
apart from the replacement $\lambda \rightarrow \tilde{\lambda}$, in
agreement with the general property (\ref{lr}): 
\begin{equation*}
\langle A_{\text{d}}^{\mu a}\hspace{0.01in}|\hspace{0.01in}A_{\text{d}}^{\nu
b}\rangle _{\text{one-loop}}=\frac{ig^{2}N_{c}(13-3\tilde{\lambda})}{6(4\pi
)^{2}(p^{2})^{2}}\delta ^{ab}(p^{2}\eta ^{\mu \nu }-p^{\mu }p^{\nu })\ln
(-p^{2}-i\epsilon ).
\end{equation*}%
As in the case of the fermion self-energy, the absorptive part, 
\begin{equation*}
\text{Abso}\left[ \langle A_{\text{d}}^{\mu a}\hspace{0.01in}|\hspace{0.01in}%
A_{\text{d}}^{\nu b}\rangle \right] =-2\text{Re}\left[ (ip^{2})\langle A_{%
\text{d}}^{\mu a}\hspace{0.01in}|\hspace{0.01in}A_{\text{d}}^{\nu b}\rangle
(ip^{2})\right] =\frac{g^{2}N_{c}(13-3\tilde{\lambda})}{48\pi }\delta
^{ab}(p^{2}\eta ^{\mu \nu }-p^{\mu }p^{\nu })\theta (p^{2}),
\end{equation*}%
is not physical, because the cloud is not physical.

Switching to purely virtual clouds, the imaginary part does not change. To
work out the real part, we just need to compute one diagram, i.e., the
self-energy diagram where physical gauge fields circulate with the
propagator (\ref{AAphys}). Note that we do not need to use the vertex $%
AA\phi $, because it involves at least one divergence $\partial ^{\mu
}A_{\mu }$, by formula (\ref{scloud2}). We obtain, for $p^{\mu }=(p^{0},%
\mathbf{0})$, 
\begin{eqnarray*}
\text{Abso}[\langle A_{\text{d}}^{ia}(p)\hspace{0.01in}A_{\text{d}%
}^{jb}(-p)\rangle ] &=&\frac{g^{2}N_{c}\delta ^{ab}\delta ^{ij}}{24\pi }%
p^{2}\theta (p^{2}), \\
\text{Abso}[\langle A_{\text{d}}^{0a}(p)\hspace{0.01in}A_{\text{d}%
}^{0b}(-p)\rangle ] &=&\text{Abso}[\langle A_{\text{d}}^{0a}(p)\hspace{0.01in%
}A_{\text{d}}^{ib}(-p)\rangle ]=0.
\end{eqnarray*}%
The results are again gauge independent, cloud independent and nonnegative.

\section{Renormalization}

\setcounter{equation}{0}\label{renormalization}

In this section we study the renormalization of the extended theory. We show
that everything goes through in the usual way (by means of renormalization
constants for the couplings, the masses, the fields and the sources) apart
from nonpolynomial, nonderivative redefinitions of the cloud fields and
their anticommuting partners into functions of themselves, with no mixings
among different cloud sectors.

\subsection{Master equations}

First, the master equation (\ref{cloudmaster}) implies an analogous master
equation%
\begin{equation}
(\Gamma _{\text{tot}},\Gamma _{\text{tot}})=0  \label{stotot}
\end{equation}%
for the generating functional $\Gamma _{\text{tot}}=W_{\text{tot}}(J,\tilde{J%
},K,\tilde{K})-\int \Phi ^{\alpha }J_{\alpha }-\sum_{i}\int \tilde{\Phi}%
^{\alpha i}\tilde{J}_{\alpha }^{i}$ of the connected, one-particle
irreducible (1PI) Green functions, where $\Phi ^{\alpha }=\delta _{r}W_{%
\text{tot}}/\delta J^{\alpha }$, $\tilde{\Phi}^{\alpha i}=\delta _{r}W_{%
\text{tot}}/\delta \tilde{J}^{\alpha i}$. The proof follows from a change of
field variables 
\begin{equation*}
\Phi ^{\alpha }\rightarrow \Phi ^{\alpha }+\theta (S_{\text{tot}},\Phi
^{\alpha }),\qquad \tilde{\Phi}^{\alpha i}\rightarrow \tilde{\Phi}^{\alpha
i}+\theta (S_{\text{tot}},\tilde{\Phi}^{\alpha i}),
\end{equation*}%
in the functional integral (\ref{Z}) that defines $Z_{\text{tot}}=\exp (iW_{%
\text{tot}})$. Only the source terms $\int \Phi J+\int \tilde{\Phi}\tilde{J}$
contribute, giving 
\begin{equation*}
\int \langle (S_{\text{tot}},\Phi ^{\alpha })\rangle J^{\alpha
}+\sum_{i}\int \langle (S_{\text{tot}},\tilde{\Phi}^{\alpha i})\rangle 
\tilde{J}^{\alpha i}=0,
\end{equation*}%
which can easily be rewritten as (\ref{stotot}). When (\ref{cloudmaster})
does not hold, the same argument gives $(\Gamma _{\text{tot}},\Gamma _{\text{%
tot}})=\langle (S_{\text{tot}},S_{\text{tot}})\rangle $.

Second, the $i$th cloud invariance of the total action $S_{\text{tot}}$,
i.e., the identity $(S_{K}^{\text{cloud\hspace{0.01in}}i},S_{\text{tot}})=0$
of (\ref{SKtoti}), implies the $i$th cloud invariance%
\begin{equation}
(S_{K}^{\text{cloud\hspace{0.01in}}i},\Gamma _{\text{tot}})=0
\label{scloudtot}
\end{equation}%
of the $\Gamma $ functional. The proof follows from the change of field
variables 
\begin{equation}
\Phi ^{\alpha }\rightarrow \Phi ^{\alpha },\qquad \tilde{\Phi}^{\alpha
i}\rightarrow \tilde{\Phi}^{\alpha i}+\theta (S_{K}^{\text{cloud\hspace{%
0.01in}}i},\tilde{\Phi}^{\alpha i}),\qquad \tilde{\Phi}^{\alpha
j}\rightarrow \tilde{\Phi}^{\alpha j}\text{ for }i\neq j\text{,}  \label{icl}
\end{equation}%
in $Z_{\text{tot}}$. Both the source terms and the action $S_{\text{tot}}$
contribute now, giving 
\begin{equation*}
\left\langle \int \frac{\delta _{r}S_{K}^{\text{cloud\hspace{0.01in}}i}}{%
\delta \tilde{K}^{\alpha i}}\frac{\delta _{l}S_{\text{tot}}}{\delta \tilde{%
\Phi}^{\alpha i}}\right\rangle =\int (S_{K}^{\text{cloud}},\tilde{\Phi}%
^{\alpha i})\tilde{J}^{\alpha i},
\end{equation*}%
which can be rewritten as (\ref{scloudtot}), after using $(S_{K}^{\text{cloud%
\hspace{0.01in}}i},S_{\text{tot}})=0$ in the left-hand side.

\subsection{Renormalization algorithm}

Proceeding inductively, we denote the order of the loop expansion by the
power of $\hbar $ (although $\hbar $ is set to one everywhere else in this
paper). We assume that we have renormalized the theory up to $n$ loops. We
denote the so-renormalized action by $S_{n\text{ tot}}$ and the $\Gamma $
functional associated with it by $\Gamma _{n\text{ tot}}$. We also assume
that $S_{n\text{ tot}}$ has the form $S_{n\text{ tot}}=S_{\text{tot}}+$
divergent counterterms (in some subtraction scheme) and satisfies 
\begin{equation*}
(S_{n\text{ tot}},S_{n\text{ tot}})=\mathcal{O}(\hbar ^{n+1}),\qquad (S_{K}^{%
\text{cloud\hspace{0.01in}}i},S_{n\text{ tot}})=0.
\end{equation*}%
The inductive assumptions are clearly satisfied at order zero.

The locality of counterterms ensures, as usual, that the order $(n+1)$
divergent part $\Gamma _{n\text{ tot div}}^{(n+1)}$ of $\Gamma _{n\text{ tot}%
}$ is local. By the argument above, the effective action $\Gamma _{n\text{
tot}}$ satisfies $(\Gamma _{n\text{ tot}},\Gamma _{n\text{ tot}})=\langle
(S_{n\text{ tot}},S_{n\text{ tot}})\rangle =(S_{n\text{ tot}},S_{n\text{ tot}%
})+\mathcal{O}(\hbar ^{n+2})$. The $(n+1)$-th order divergent part of this
equation gives 
\begin{equation*}
2(S_{\text{tot}},\Gamma _{n\text{ tot div}}^{(n+1)})=(S_{n\text{ tot}},S_{n%
\text{ tot}})+\mathcal{O}(\hbar ^{n+2}),
\end{equation*}%
having noted that $(S_{n\text{ tot}},S_{n\text{ tot}})$ is divergent.
Defining%
\begin{equation}
S_{n+1\text{ tot}}=S_{n\text{ tot}}-\Gamma _{n\text{ tot div}}^{(n+1)}=S_{%
\text{tot}}+\text{divergent counterterms},  \label{subtran}
\end{equation}%
we have 
\begin{equation*}
(S_{n+1\text{ tot}},S_{n+1\text{ tot}})=(S_{n\text{ tot}},S_{n\text{ tot}%
})-2(S_{n\text{ tot}},\Gamma _{n\text{ tot div}}^{(n+1)})+\mathcal{O}(\hbar
^{n+2})=\mathcal{O}(\hbar ^{n+2}).
\end{equation*}%
Moreover, $(S_{K}^{\text{cloud\hspace{0.01in}}i},S_{n\text{ tot}})=0$
implies $(S_{K}^{\text{cloud\hspace{0.01in}}i},\Gamma _{n\text{ tot}})=0$
and its $(n+1)$-th order divergent part gives $(S_{K}^{\text{cloud\hspace{%
0.01in}}i},\Gamma _{n\text{ tot div}}^{(n+1)})=0$, which in turn implies $%
(S_{K}^{\text{cloud\hspace{0.01in}}i},S_{n+1\text{ tot}})=0$. Finally, by (%
\ref{subtran}) $\Gamma _{n+1\text{ tot}}$ is convergent up to the order $n+1$
included. Thus, the inductive assumptions are fully replicated to that
order. This allows us to take the argument to $n\rightarrow \infty $, where
we obtain the renormalized action $S_{R\hspace{0.01in}\text{tot}}\equiv
S_{\infty \text{ tot}}$, and conclude that it satisfies the renormalized
master equations 
\begin{equation}
(S_{R\hspace{0.01in}\text{tot}},S_{R\hspace{0.01in}\text{tot}})=0,\qquad
(S_{K}^{\text{cloud\hspace{0.01in}}i},S_{R\hspace{0.01in}\text{tot}})=0.
\label{SRtotmast}
\end{equation}

\subsection{Renormalized action}

Now we characterize $S_{R\hspace{0.01in}\text{tot}}$ more precisely. Besides
the usual ghost number, we introduce \textquotedblleft cloud
numbers\textquotedblright\ for each cloud. The usual ghost number is equal
to 1 for $C$, minus 1 for $\bar{C}$, $K_{B}$, $K_{A}^{\mu }$, ${K_{\psi }}$, 
${K_{\bar{\psi}}}$, $\tilde{K}_{\phi }^{i}$ and $\tilde{K}_{H}^{i}$, minus 2
for $K_{C}$, and 0 for every other field and source. The $i$th cloud number
is equal to one for $H^{i}$, minus one for $\bar{H}^{i}$, $\tilde{K}_{H}^{i}$
and $\tilde{K}_{E}^{i}$, and zero in all the other cases.

Every term of the action $S_{\text{tot}}$ is neutral with respect to the
ghost and cloud numbers just defined, with the exception of the source terms 
$\int H^{i}\tilde{K}_{\phi }^{i}$. Since, however, such terms cannot be used
in nontrivial 1PI diagrams, all the counterterms are neutral. Thus, each
cloud number is separately conserved by the 1PI diagrams beyond the tree
level.

By power counting, the counterterms can be at most linear in the sources.
Indeed, the dimensions of $\tilde{K}_{\phi }^{i}$, $\tilde{K}_{H}^{i}$, $%
\tilde{K}_{\bar{H}}^{i}$, $\tilde{K}_{E}^{i}$ are 3, 2, 2 and 1,
respectively, but $\tilde{K}_{E}^{i}$ never appears, $\tilde{K}_{\bar{H}%
}^{i} $ appears trivially and does not participate in the counterterms,
while a bilinear in $\tilde{K}_{H}^{i}$ is prohibited by the conservation of
the $i$th cloud number.

We recall that the cloud symmetry generated by $S_{K}^{\text{cloud}}$
collects the most general shifts of the cloud fields $\phi ^{i}$, combined
with analogous shifts of $\bar{H}^{i}$ and the sources $\tilde{K}_{H}^{i}$
and $\tilde{K}_{E}^{i}$. A general theorem (which is easily proved by
switching to the language of differential forms -- see, for example, the
appendix of \cite{mastercanc}) ensures that a local functional $X$ that is
closed with respect to a symmetry of this type (i.e., such that $(S_{K}^{%
\text{cloud}},X)=0$) is the sum of an exact local functional (i.e., a
functional of the form $(S_{K}^{\text{cloud}},Y)$, for some other local
functional $Y$) plus a local functional that is independent on the shifted
fields, as well as their shifts.

Since $S_{R\hspace{0.01in}\text{tot}}$ satisfies the second equation (\ref%
{SRtotmast}) for every $i$, hence $(S_{K}^{\text{cloud}},S_{R\hspace{0.01in}%
\text{tot}})=0$, it can be written as the sum%
\begin{equation}
S_{R\hspace{0.01in}\text{tot}}=S_{R}+(S_{K}^{\text{cloud\hspace{0.01in}}%
},\Upsilon _{R})+S_{K}^{\text{cloud}}  \label{StotRR}
\end{equation}%
of a local functional $S_{R}$ that does not depend on the cloud fields and
the cloud sources, plus a cloud exact functional, where $\Upsilon _{R}$ is
local. We have separated $S_{K}^{\text{cloud}}$ from the rest, because $%
S_{K}^{\text{cloud}}$ is nonrenormalized, due to its triviality.

\subsection{Cloud independence through renormalization}

\label{cloudindepr}

Now we prove that the functional $S_{R}$ coincides with the usual
renormalized action. To achieve this goal, we need to show that the cloud
independence theorem of section \ref{cloudindep} safely goes through the
renormalization algorithm. The proof given in section \ref{cloudindep},
which relies on the specific forms of the action and the cloud fermions used
there, needs to be upgraded in a nontrivial way.

Since every renormalized action $S_{n\text{ tot}}$, as well as the
functionals $\Gamma _{n\text{ tot div}}^{(n+1)}$, satisfy $(S_{K}^{\text{%
cloud}},S_{n\text{ tot}})=(S_{K}^{\text{cloud}},\Gamma _{n\text{ tot div}%
}^{(n+1)})=0$, the argument used for (\ref{StotRR}) allows us to write them
as 
\begin{equation}
S_{n\text{ tot}}=S_{n}+(S_{K}^{\text{cloud}},\Upsilon _{n})+S_{K}^{\text{%
cloud}},\qquad \Gamma _{n\text{ tot div}}^{(n+1)}=\Gamma _{n\text{ div}%
}^{(n+1)}+(S_{K}^{\text{cloud}},R_{n\text{ div}}^{(n+1)}),  \label{due}
\end{equation}%
where $S_{n}$ and $\Gamma _{n\text{ div}}^{(n+1)}$ are independent of the
cloud fields $\tilde{\Phi}^{\alpha i}$ and the cloud sources $\tilde{K}%
^{\alpha i}$.

Assume, by induction, that $S_{n}$ is cloud independent (that is to say,
independent of the cloud parameters $\tilde{\lambda}$). Let $Z_{n\hspace{%
0.01in}\text{tot}}(J,K,\tilde{J},\tilde{K})$ denote the generating
functional associated with the action $S_{n\hspace{0.01in}\text{tot}}$. At $%
\tilde{J}=\tilde{K}=0$ it reads 
\begin{eqnarray}
Z_{n\hspace{0.01in}\text{tot}}(J,K,0,0) &=&\int [\mathrm{d}\Phi \mathrm{d}%
\tilde{\Phi}]\mathrm{\exp }\left( iS_{n}(\Phi ,K)+i(S_{K}^{\text{cloud%
\hspace{0.01in}}},\bar{\Upsilon}_{n})+i\int \Phi ^{\alpha }J_{\alpha }\right)
\notag \\
&=&\int [\mathrm{d}\Phi ]\mathrm{\exp }\left( iS_{n}(\Phi ,K)+i\int \Phi
^{\alpha }J_{\alpha }\right) \int [\mathrm{d}\tilde{\Phi}]\mathrm{e}%
^{i(S_{K}^{\text{cloud}},\bar{\Upsilon}_{n})},  \label{Ztot}
\end{eqnarray}%
where $\bar{\Upsilon}_{n}$ denotes $\Upsilon _{n}$ at $\tilde{K}=0$. The
cloud sector does not contribute, because%
\begin{equation}
\int [\mathrm{d}\tilde{\Phi}]\mathrm{e}^{i(S_{K}^{\text{cloud}},\bar{\Upsilon%
}_{n})}=1.  \label{ident}
\end{equation}%
This identity can be proved as follows. The left-hand side is in principle a
functional of the fields $\Phi $ and the sources $K$, since $\bar{\Upsilon}%
_{n}$ may depend on them. To show that it is actually a constant, we
consider arbitrary infinitesimal variations of $\Phi $ and $K$. If $\delta 
\bar{\Upsilon}_{n}$ denotes the variation of $\bar{\Upsilon}_{n}$ due to
them, the variation of the integral is 
\begin{equation}
\delta \int [\mathrm{d}\tilde{\Phi}]\mathrm{e}^{i(S_{K}^{\text{cloud}},\bar{%
\Upsilon}_{n})}=i\int [\mathrm{d}\tilde{\Phi}](S_{K}^{\text{cloud}},\delta 
\bar{\Upsilon}_{n})\mathrm{e}^{i(S_{K}^{\text{cloud}},\bar{\Upsilon}_{n})},
\label{varia}
\end{equation}%
Performing the change of field variables $\tilde{\Phi}^{\alpha }\rightarrow 
\tilde{\Phi}^{\alpha }+\theta (S_{K}^{\text{cloud}},\tilde{\Phi}^{\alpha })$
in the integral%
\begin{equation*}
\int [\mathrm{d}\tilde{\Phi}]\hspace{0.01in}\delta \bar{\Upsilon}_{R}\hspace{%
0.01in}\mathrm{e}^{i(S_{K}^{\text{cloud}},\bar{\Upsilon}_{n})},
\end{equation*}%
we obtain%
\begin{equation}
\int [\mathrm{d}\tilde{\Phi}]\hspace{0.01in}\delta \bar{\Upsilon}_{n}\hspace{%
0.01in}\mathrm{e}^{i(S_{K}^{\text{cloud}},\bar{\Upsilon}_{n})}=\int [\mathrm{%
d}\tilde{\Phi}]\left[ \delta \bar{\Upsilon}_{n}+\theta (S_{K}^{\text{cloud}%
},\delta \bar{\Upsilon}_{n})\right] \mathrm{e}^{i(S_{K}^{\text{cloud}},\bar{%
\Upsilon}_{n})}.  \label{cloudin}
\end{equation}%
We have used the fact that $(S_{K}^{\text{cloud}},\bar{\Upsilon}_{n})$ is
independent of the sources $\tilde{K}$, so $(S_{K}^{\text{cloud}},\bar{%
\Upsilon}_{n})\rightarrow (S_{K}^{\text{cloud}},\bar{\Upsilon}_{n})+\theta
(S_{K}^{\text{cloud}},(S_{K}^{\text{cloud}},\bar{\Upsilon}_{n}))=(S_{K}^{%
\text{cloud}},\bar{\Upsilon}_{n})$. The equality (\ref{cloudin}) shows that
the right-hand side of (\ref{varia}) vanishes, as we wished to prove.

Thus, all the connected correlation functions of the undressed fields, which
are collected in $W_{n\hspace{0.01in}\text{tot}}(J,K,0,0)$, coincide with
the usual ones, even at the renormalized level. Not only, we can also show
that the 1PI correlation functions of the undressed fields, collected in $%
\Gamma _{n\hspace{0.01in}\text{tot}}(\Phi ,K,0,0)$, coincide with the usual
ones. Indeed, it is easy to see, using the second equation of (\ref{due}),
that setting $\tilde{\Phi}^{i}=0$ is equivalent to setting $\tilde{J}^{i}=0$
in all cases apart from $\tilde{J}_{E}^{i}$. The proof given above also
works if we keep the sources $\tilde{J}_{E}^{i}$ arbitrary, since $%
\sum_{i}\int E^{i}\tilde{J}_{E}^{i}=(S_{K}^{\text{cloud}},\sum_{i}\int \bar{H%
}^{i}\tilde{J}_{E}^{i})$: the derivation can be repeated with $\bar{\Upsilon}%
_{n}\rightarrow \bar{\Upsilon}_{n}+\sum_{i}\int \bar{H}^{i}\tilde{J}_{E}^{i}$%
. Thus, even $W_{n\hspace{0.01in}\text{tot}}(J,K,\hat{0},0)$ coincides with
the usual one, where $\hat{0}$ means that all the sources $\tilde{J}^{i}$
are set to zero but $\tilde{J}_{E}^{i}$. Actually, $W_{n\hspace{0.01in}\text{%
tot}}(J,K,\hat{0},0)$ does not even depend on $\tilde{J}_{E}^{i}$. These
facts imply that $\Gamma _{n\hspace{0.01in}\text{tot}}(\Phi ,K,0,0)$
coincides with the usual non-cloud one, and so does $\Gamma _{n\text{ div}%
}^{(n+1)}$. In particular, $\Gamma _{n\text{ div}}^{(n+1)}$ is cloud
independent. Then, (\ref{subtran}) shows that $S_{n+1}$, inside $S_{n+1\text{%
\hspace{0.01in}tot}}$, is cloud independent. Finally, we can take $n$ to
infinity, and infer that $S_{R}$ is cloud independent and coincides with the
usual renormalized action.

We have thus achieved a neat separation between the fundamental theory and
the cloud sectors, and ensured that the separation is compatible with
renormalization. We recall that $S_{R}$ is determined by gauge-independent
renormalization constants $Z_{g}$ and $Z_{m}$ for the coupling $g$ and the
fermion mass $m$, respectively, plus a generically gauge-dependent canonical
transformation that incorporates the wave-function renormalization constants
of the fields $\Phi $ and the sources $K$.

The arguments of this subsection can be specialized to every cloud, to prove
that the $i$th cloud parameters $\tilde{\lambda}_{i}$ do not propagate to
the other cloud sectors.

\subsection{Renormalized clouds}

Now we analyze the functional $\Upsilon _{R}$. Separating the
source-independent part $\tilde{\Psi}_{R}$ of $\Upsilon _{R}$ from the
source-dependent part, we can write 
\begin{equation}
(S_{K}^{\text{cloud}},\Upsilon _{R})=(S_{K}^{\text{cloud}},\tilde{\Psi}%
_{R})+\sum_{i}\int (C^{a}S^{abi}(\phi )\tilde{K}_{\phi
}^{bi}+C^{a}H^{bi}S^{abci}(\phi )\tilde{K}_{H}^{ci}),  \label{supsi}
\end{equation}%
for some functions $S^{abi}(\phi )$ and $S^{abci}(\phi )$, which encode the
gauge transformations of the cloud fields and the cloud ghosts. The
structures of the last two terms are determined by the conservations of the
ghost and cloud numbers, as well as power counting and cloud exactness.
These same properties exclude any other source-dependent terms.

The functions $S^{abi}(\phi )$ and $S^{abci}(\phi )$ are not independent,
since by cloud exactness it must be possible to collect the last two terms
of (\ref{supsi}) into%
\begin{equation}
\left( S_{K}^{\text{cloud}},\sum_{i}\int C^{a}S^{abi}(\phi )\tilde{K}%
_{H}^{bi}\right) .  \label{imply}
\end{equation}%
Moreover, each function $S^{abi}(\phi )$ can depend only on the $i$th cloud
field $\phi ^{i}$, because otherwise (\ref{imply}) is not neutral with
respect to each cloud number separately. Thus, from now on we write $%
S^{abi}(\phi ^{i})$.

Consider $(S_{K}^{\text{cloud}},\tilde{\Psi}_{R})$. The gauge and cloud
conditions we have used in this paper, which are (\ref{covg}), (\ref{cloud}%
), (\ref{spepsi}) and (\ref{spepsit}), ensure that the counterterms can
depend on $\bar{C}$, $\bar{H}^{i}$ and $E^{i}$ only through the derivatives $%
\partial _{\mu }\bar{C}$, $\partial _{\mu }\bar{H}^{i}$, $\partial _{\mu
}E^{i}$. Moreover, they cannot depend on $B$. We want to show that the
renormalized gauge fermion $\tilde{\Psi}_{R}$ has the form 
\begin{equation}
\tilde{\Psi}_{R}=\sum_{i}\int \bar{H}^{ai}\left( V_{R}^{ai}(A,\phi ^{i})+%
\frac{\tilde{\lambda}_{i}}{2}E^{ai}\right)  \label{psitR}
\end{equation}%
for some local functions $V_{R}^{ai}$ that depend only on the $i$th cloud
fields $\phi ^{i}$. First, note that a term proportional to $\bar{C}$ cannot
appear in $\tilde{\Psi}_{R}$, because its antiparenthesis with $S_{K}^{\text{%
cloud}}$ would have dimension greater than four, or not be neutral with
respect to the ghost and cloud numbers.

Second, the coefficient of $\bar{H}^{ai}$ in $\tilde{\Psi}_{R}$ has
dimension 2. It cannot contain $B$, because $B$ appears trivially in the
action. It cannot depend on $\bar{C}$ and $C$ either, because $(S_{K}^{\text{%
cloud}},\tilde{\Psi}_{R})$ would contain a term $\sim E\bar{C}C$, which
cannot be generated, since $E$ and $\bar{C}$ can appear only through their
derivatives. For the same reason, the terms $\sim \bar{H}E$ of $\tilde{\Psi}%
_{R}$ are nonrenormalized, because any corrections would bring counterterms $%
\sim E^{2}$ in $(S_{K}^{\text{cloud}},\tilde{\Psi}_{R})$.

Thus, $\tilde{\Psi}_{R}$ can only contain the gauge fields and the cloud
fields in the way we have shown in (\ref{psitR}). Moreover, the functions $%
V_{R}^{ai}$ can only depend on the $i$th cloud field $\phi ^{i}$, otherwise $%
(S_{K}^{\text{cloud}},\tilde{\Psi}_{R})$ would violate the $i$th cloud
number conservation. This proves that there are no mixings among different
cloud sectors.

Finally, the renormalized action reads%
\begin{equation}
S_{R\hspace{0.01in}\text{tot}}=S_{R}+\sum_{i}\left( S_{K}^{\text{cloud%
\hspace{0.01in}}i},\int \bar{H}^{ai}\left( V_{R}^{ai}(A,\phi ^{i})+\frac{%
\tilde{\lambda}_{i}}{2}E^{ai}\right) +\int C^{a}S^{abi}(\phi ^{i})\tilde{K}%
_{H}^{bi}\right) +S_{K}^{\text{cloud}},\qquad  \label{StotR3}
\end{equation}%
It may also be convenient to organize it as%
\begin{equation}
S_{R\hspace{0.01in}\text{tot}}=S_{R\hspace{0.01in}\text{tot\hspace{0.01in}}%
0}+S_{K,R}^{\text{gauge}}+S_{K}^{\text{cloud}},  \label{SRtot4}
\end{equation}%
where $S_{R\hspace{0.01in}\text{tot\hspace{0.01in}}0}=\left. S_{R\hspace{%
0.01in}\text{tot}}\right\vert _{K=\tilde{K}=0}$. The renormalized gauge
transformations are encoded in $S_{K,R}^{\text{gauge}}$. Separating the
various contributions according to their dependences on the sources $K$ and $%
\tilde{K}$, the master equations (\ref{SRtotmast}) imply%
\begin{eqnarray*}
(S_{K,R}^{\text{gauge}}+S_{K}^{\text{cloud}},S_{R\hspace{0.01in}\text{tot%
\hspace{0.01in}}0}) &=&0,\qquad (S_{K,R}^{\text{gauge}}+S_{K}^{\text{cloud}%
},S_{K,R}^{\text{gauge}}+S_{K}^{\text{cloud}})=0, \\
(S_{K}^{\text{cloud\hspace{0.01in}}i},S_{R\hspace{0.01in}\text{tot\hspace{%
0.01in}}0}) &=&(S_{K}^{\text{cloud\hspace{0.01in}}i},S_{K,R}^{\text{gauge}%
}+S_{K}^{\text{cloud}})=0,
\end{eqnarray*}%
which immediately give%
\begin{equation}
(S_{K,R}^{\text{gauge}},S_{R\hspace{0.01in}\text{tot\hspace{0.01in}}%
0})=(S_{K}^{\text{cloud\hspace{0.01in}}i},S_{R\hspace{0.01in}\text{tot%
\hspace{0.01in}}0})=0,\qquad (S_{K,R}^{\text{gauge}},S_{K,R}^{\text{gauge}%
})=(S_{K}^{\text{cloud\hspace{0.01in}}i},S_{K,R}^{\text{gauge}})=0.
\label{renogauge}
\end{equation}

Combined, the two sets of equations imply $(S_{K,R}^{\text{gauge}},S_{R%
\hspace{0.01in}\text{tot}})=0$, which was not obvious from (\ref{SRtotmast}%
). The left equations ensure that $S_{R\hspace{0.01in}\text{tot\hspace{0.01in%
}}0}$ is gauge and cloud invariant. In particular, all the functions $%
V_{R}^{ai}$ of (\ref{StotR3}) must be gauge invariant, by the gauge
invariance of the terms $E^{ai}V_{R}^{ai}$ contained in $(S_{K}^{\text{cloud}%
},\tilde{\Psi}_{R})$.

The functions $S^{abi}(\phi ^{i})$ are further constrained by the closure of
the renormalized gauge transformations. Apart from that, they are arbitrary.
Indeed, it is always possible to make nontrivial redefinitions that send
each $\phi ^{i}$ into a function of itself. Then, for consistency, each $%
H^{i}$ must be sent into $H^{i}$ times a suitable function of $\phi ^{i}$.
Since the fields $\phi ^{i}$ are dimensionless, renormalization can activate
nonpolynomial, nonderivative redefinitions of this type. By the second
equation of (\ref{cind}), these redefinitions can depend on the gauge-fixing
parameters and the $i$th cloud parameters, but not on the parameters of the
other clouds.

\subsection{Renormalized dressed fields}

Using the renormalized gauge transformations, which are encoded in the
functional $S_{K,R}^{\text{gauge}}$, we can build the renormalized, gauge
invariant dressed fields $A_{\mu \text{d}R}$, $\psi _{\text{d}R}$ and $\bar{%
\psi}_{\text{d}R}$. By the result of appendix \ref{uniqueness}, their
expressions are unique up to constant (matrix) factors. We want to prove
that we can fix those factors so that the correlation functions of $A_{\mu 
\text{d}R}$, $\psi _{\text{d}R}$ and $\bar{\psi}_{\text{d}R}$ are gauge
independent.

Including appropriate sources, the extended renormalized action is%
\begin{equation*}
S_{R\hspace{0.01in}\text{tot}}^{\text{ext}}=S_{R\hspace{0.01in}\text{tot}%
}+\int \left( J_{\text{d}R}^{\mu }A_{\mu \text{d}R}+\bar{J}_{\psi \text{d}%
R}\psi _{\text{d}R}+\bar{\psi}_{\text{d}R}J_{\psi \text{d}R}\right) .
\end{equation*}%
Clearly, $(S_{K,R}^{\text{gauge}},S_{R\hspace{0.01in}\text{tot\hspace{0.01in}%
}0}^{\text{ext}})=0$, where $S_{R\hspace{0.01in}\text{tot\hspace{0.01in}}0}^{%
\text{ext}}=\left. S_{R\hspace{0.01in}\text{tot}}^{\text{ext}}\right\vert
_{K=\tilde{K}=0}$.

The equations of gauge dependence (\ref{gind}), derived in appendix \ref%
{appe3}, evaluated at $K=\tilde{K}=0$, give%
\begin{eqnarray}
\frac{\partial S_{R\hspace{0.01in}\text{tot\hspace{0.01in}}0}}{\partial
\lambda } &=&(S_{R\hspace{0.01in}\text{tot\hspace{0.01in}}0},\Psi _{\lambda 
\hspace{0.01in}R\hspace{0.01in}1})+(S_{K,R}^{\text{gauge}},\Psi _{\lambda 
\hspace{0.01in}R\hspace{0.01in}0}),  \notag \\
\frac{\partial S_{K,R}^{\text{gauge}}}{\partial \lambda } &=&(S_{K,R}^{\text{%
gauge}},\Psi _{\lambda \hspace{0.01in}R\hspace{0.01in}1}),\qquad (S_{K}^{%
\text{cloud\hspace{0.01in}}i},\Psi _{\lambda \hspace{0.01in}R\hspace{0.01in}%
0})=(S_{K}^{\text{cloud\hspace{0.01in}}i},\Psi _{\lambda \hspace{0.01in}R%
\hspace{0.01in}1})=0,  \label{eques}
\end{eqnarray}%
where $\Psi _{\lambda \hspace{0.01in}R\hspace{0.01in}0}=\left. \Psi
_{\lambda \hspace{0.01in}R}\right\vert _{K=\tilde{K}=0}$ and $\Psi _{\lambda 
\hspace{0.01in}R\hspace{0.01in}1}=\Psi _{\lambda \hspace{0.01in}R}-\Psi
_{\lambda \hspace{0.01in}R\hspace{0.01in}0}$. In particular, the first
formula gives%
\begin{equation*}
\frac{\partial S_{R\hspace{0.01in}\text{tot\hspace{0.01in}}0}}{\partial
\lambda }=\int \frac{\delta _{r}S_{R\hspace{0.01in}\text{tot\hspace{0.01in}}%
0}}{\delta \hat{\Phi}^{\alpha }}\Delta \hat{\Phi}^{\alpha }+(S_{K,R}^{\text{%
gauge}},\Psi _{\lambda \hspace{0.01in}R\hspace{0.01in}0}),\qquad \Delta \hat{%
\Phi}^{\alpha }=\frac{\delta _{l}\Psi _{\lambda \hspace{0.01in}R\hspace{%
0.01in}1}}{\delta \hat{K}^{\alpha }},
\end{equation*}%
where $\hat{\Phi}^{\alpha }=(\Phi ^{\alpha },\tilde{\Phi}^{\alpha })$ and $%
\hat{K}^{\alpha }=(K^{\alpha },\tilde{K}^{\alpha })$. This result tells us
the the whole gauge dependence of $S_{R\hspace{0.01in}\text{tot\hspace{0.01in%
}}0}$ is encoded into a field redefinition, plus a gauge-exact term. The
field redefinition $\hat{\Phi}^{\alpha }=\hat{\Phi}^{\alpha }(\hat{\Phi}%
^{\prime },\lambda )$ is the solution of%
\begin{equation*}
\frac{\partial \hat{\Phi}^{\alpha }(\hat{\Phi}^{\prime },\lambda )}{\partial
\lambda }=-\Delta \hat{\Phi}^{\alpha }(\hat{\Phi}(\hat{\Phi}^{\prime
},\lambda ),\lambda ),
\end{equation*}%
with arbitrary initial conditions. It can be worked out perturbatively in $g$%
, starting from $\hat{\Phi}^{\alpha }=\hat{\Phi}^{\alpha \hspace{0.01in}%
\prime }$. \ 

It is convenient to switch to the new variables $\hat{\Phi}^{\alpha \hspace{%
0.01in}\prime }$, $\hat{K}^{\alpha \hspace{0.01in}\prime }$ by means of the
canonical transformation generated by $F(\hat{\Phi},\hat{K}^{\prime })=\int 
\hat{\Phi}^{\prime }(\hat{\Phi},\lambda )\hat{K}^{\prime }$. In so doing, we
obtain 
\begin{equation}
\frac{\partial S_{R\hspace{0.01in}\text{tot\hspace{0.01in}}0}^{\prime }}{%
\partial \lambda }=(S_{K,R}^{\text{gauge\hspace{0.01in}}\prime },\Psi
_{\lambda \hspace{0.01in}R\hspace{0.01in}0}^{\prime })^{\prime },\qquad 
\frac{\partial S_{K,R}^{\text{gauge\hspace{0.01in}}\prime }}{\partial
\lambda }=0,  \label{primed}
\end{equation}%
where the prime on the antiparentheses refers to the new variables. The last
equation follows from 
\begin{equation*}
\frac{\partial \hat{K}^{\alpha }(\hat{\Phi}^{\prime },\hat{K}^{\prime
},\lambda )}{\partial \lambda }=\frac{\delta _{l}\Psi _{\lambda \hspace{%
0.01in}R\hspace{0.01in}1}}{\delta \hat{\Phi}^{\alpha }}(\hat{\Phi}(\hat{\Phi}%
^{\prime }),\hat{K}(\hat{\Phi}^{\prime },\hat{K}^{\prime })),
\end{equation*}%
which is easy to prove from the transformation.

Now we consider the correlations functions that contain insertions of $%
A_{\mu \text{d}R}$, $\psi _{\text{d}R}$ and $\bar{\psi}_{\text{d}R}$, and
apply arguments that are analogous to those of subsection \ref{gaugeproof}.
First, we switch to the variables with primes everywhere. We denote the
transformed action $S_{R\hspace{0.01in}\text{tot}}(\hat{\Phi}(\hat{\Phi}%
^{\prime }),\hat{K}(\hat{\Phi}^{\prime },$ $\hat{K}^{\prime }))$ by $S_{R%
\hspace{0.01in}\text{tot}}^{\prime }$ and the transformed renormalized
dressed fields $A_{\mu \text{d}R}(\hat{\Phi}(\hat{\Phi}^{\prime }))$, $\psi
_{\text{d}R}(\hat{\Phi}(\hat{\Phi}^{\prime }))$ and $\bar{\psi}_{\text{d}R}(%
\hat{\Phi}(\hat{\Phi}^{\prime }))$ by $A_{\mu \text{d}R}^{\prime }$, $\psi _{%
\text{d}R}^{\prime }$ and $\bar{\psi}_{\text{d}R}^{\prime }$. Gauge
invariance, which reads $(S_{K,R}^{\text{gauge}},S_{R\hspace{0.01in}\text{tot%
\hspace{0.01in}}0}^{\text{ext}})=0$, is obviously preserved by the
transformation: $(S_{K,R}^{\text{gauge\hspace{0.01in}}\prime },S_{R\hspace{%
0.01in}\text{tot\hspace{0.01in}}0}^{\text{ext\hspace{0.01in}}\prime
})^{\prime }=0$. The transformed fields $A_{\mu \text{d}R}^{\prime }$, $\psi
_{\text{d}R}^{\prime }$ and $\bar{\psi}_{\text{d}R}^{\prime }$ are $S_{K,R}^{%
\text{gauge\hspace{0.01in}}\prime }$-closed, i.e., solutions $X^{\prime }$
of $(S_{K,R}^{\text{gauge\hspace{0.01in}}\prime },X^{\prime })^{\prime }=0$,
using variables with primes.

This is where we fix the arbitrary constant factors in front of such
solutions: it is sufficient to require that $A_{\mu \text{d}R}^{\prime }$, $%
\psi _{\text{d}R}^{\prime }$ and $\bar{\psi}_{\text{d}R}^{\prime }$ be gauge
independent. Such a requirement does make sense, because the second formula
of (\ref{primed}) ensures that the gauge transformations themselves are
gauge independent in the new variables. We just have to pay attention that
the overall factors of the solutions do not introduce spurious gauge
dependencies. Once this is done, we are ready to repeat the arguments of
subsection \ref{gaugeproof}, with the replacements%
\begin{equation*}
S_{\text{tot}}^{\text{ext}}\rightarrow S_{R\hspace{0.01in}\text{tot}%
}^{\prime }+\int \left( J_{\text{d}R}^{\mu }A_{\mu \text{d}R}^{\prime }+\bar{%
J}_{\psi \text{d}R}\psi _{\text{d}R}^{\prime }+\bar{\psi}_{\text{d}%
R}^{\prime }J_{\psi \text{d}R}\right) ,\qquad \Psi _{\lambda }\rightarrow
\Psi _{\lambda \hspace{0.01in}R\hspace{0.01in}0}^{\prime },\qquad S_{K}^{%
\text{gauge}}\rightarrow S_{K,R}^{\text{gauge\hspace{0.01in}}\prime },
\end{equation*}%
besides of course $\hat{\Phi}$, $\hat{K}\rightarrow \hat{\Phi}^{\prime }$, $%
\hat{K}^{\prime }$. The result is that the correlation functions of the
renormalized dressed fields $A_{\mu \text{d}R}$, $\psi _{\text{d}R}$ and $%
\bar{\psi}_{\text{d}R}$ are gauge independent.

The renormalized sources $J_{\text{d}R}^{\mu }$, $J_{\psi \text{d}R}$ and $%
\bar{J}_{\psi \text{d}R}$ are equal to $J_{\text{d}}^{\mu }$, $J_{\psi \text{%
d}}$ and $\bar{J}_{\psi \text{d}}$ times suitable renormalization constants.
Apart from that, the correlation functions of $A_{\mu \text{d}R}$, $\psi _{%
\text{d}R}$ and $\bar{\psi}_{\text{d}R}$ do not need further
renormalization. Indeed, the sources $J_{\text{d}R}^{\mu }$, $J_{\psi \text{d%
}R}$ and $\bar{J}_{\psi \text{d}R}$ have dimensions 3, 5/2 and 5/2,
respectively, so no local counterterms with two or more of them are allowed.

Finally, the arguments that lead to the identity (\ref{onshecorr}) continue
to hold after renormalization. We obtain an identity analogous to (\ref%
{onshecorr}), where the dressed and undressed fields are replaced by their
renormalized versions. In particular, the $S$-matrix amplitudes of the
renormalized dressed fields are cloud independent, and coincide with the
usual amplitudes of the renormalized undressed fields. Since the former are
manifestly gauge independent, the latter are gauge independent as well.

\subsection{Renormalization recap}

Summarizing, the renormalized action has the structure of the starting
action, with standard multiplicative renormalization constants for the
coupling and the masses, combined with a canonical transformation that
encodes multiplicative renormalizations of the sources and the fields,
except for the cloud fields $\phi ^{i}$ and their anticommuting partners $%
H^{i}$, which are renormalized in nonpolynomial, nonderivative ways.
Different cloud sectors to not mix with one another.

From a specific gauge, every other gauge can be reached by means of a
canonical transformation. Thus, the renormalization in every other gauge is
the same as above, up to a renormalized canonical transformation. Finally,
every cloud choices can be reached from specific cloud choices by means of
canonical transformations. Again, the renormalization is the same as above
up to renormalized canonical transformations.

Equipped with the renormalized action and the renormalized gauge
transformations, we can build dressed fields that are gauge-invariant with
respect to the latter. They are unique up to constant factors, by the
theorem proved in appendix \ref{uniqueness}. The constant factors can be
fixed so that their correlation functions are gauge independent.

The proof of the gauge/cloud duality can be repeated for the renormalized
theory.

\section{Comparison with other approaches}

\setcounter{equation}{0}\label{comparison}

In this section we compare the approach of this paper with other approaches
that are available in the literature.

In the Dirac approach \cite{Dirac} the gauge invariant dressings of
electrons in QED\ are defined by means of nonlocal operator insertions, such
as%
\begin{equation}
\exp \left( -ie\frac{\nabla \cdot \mathbf{A}}{\triangle }\right) \psi ,
\label{Diracpsi}
\end{equation}%
where $\triangle $ denotes the Laplacian. We may call the exponential
prefactor \textquotedblleft Coulomb-Dirac cloud\textquotedblright . Photons
do not need a particular dressing, since we can work directly with the field
strength, which is linear in the gauge field.

The extension of the Dirac approach to quarks and non-Abelian gauge fields
has been done by Lavelle and McMullan in refs. \cite{Lavelle}. Static quarks
are dressed similarly to (\ref{Diracpsi}), while moving quarks are described
by means of boosted Coulomb-Dirac clouds. The renormalization is studied in 
\cite{Bagan}.

In the static case, the Lavelle-McMullan expressions of the dressed gauge
fields and fermions, and their correlation functions, are related to the
ones defined here as follows. First, choose a cloud function $V^{a}(A,\phi )$
of the Coulomb-Dirac type and take $\tilde{\lambda}=0$:%
\begin{equation}
\tilde{\Psi}(\Phi ,\tilde{\Phi})=\int \bar{H}^{a}(\mathbf{\nabla }\cdot 
\mathbf{A}_{\text{d}}^{a}),\qquad (S_{K}^{\text{cloud}},\tilde{\Psi})=\int
E^{a}(\mathbf{\nabla }\cdot \mathbf{A}_{\text{d}}^{a})-\int \bar{H}^{a}%
\hspace{0.01in}\mathbf{\nabla }\cdot \frac{\partial \mathbf{A}_{\text{d}%
}^{a}(A,\phi )}{\partial \phi ^{b}}H^{b}.  \label{sclou}
\end{equation}%
Then integrate $E^{a}$ out. This gives the cloud condition 
\begin{equation}
0=\mathbf{\nabla }\cdot \mathbf{A}_{\text{d}}^{a}(A,\phi )=\mathbf{\nabla }%
\cdot \mathbf{A}^{a}+\triangle \phi ^{a}+\mathcal{O}(g),  \label{Lavecod}
\end{equation}%
which can be solved perturbatively for $\phi $. The solution $\phi (\mathbf{A%
})$ is nonlocal in space, but unambiguous (and does not need a particular
prescription, since it is of the Coulomb type). If we insert it in the
expressions (\ref{dressedfields}) of $A_{\mu \text{d}}$ and $\psi _{\text{d}%
} $, the Lavelle-McMullan correlation functions are the correlation
functions of such dressed fields. Note that after these operations the cloud
sector of the action can be dropped, since it integrates to one, as in (\ref%
{ide}).

The comparison in the nonstatic case is less straightforward. Without a
general notion of pure virtuality, like the one provided by the fakeon
prescription, clouds of the Coulomb-Dirac and Lavelle-McMullan types seem to
be the meaningful choices.

Another approach to build gauge invariant correlation functions is the one
suggested by 't Hooft in ref. \cite{tHooftcloud}. Consider fermions $\psi $
in the fundamental representation and introduce scalar fields $\phi $, also
in the fundamental representation. The bilinear $\phi ^{\dagger }\psi $ is
obviously a color singlet. If a spontaneous symmetry breaking mechanism
gives $\phi $ an expectation value $v$, then the product $\phi ^{\dagger
}\psi $ can be expanded, and its expansion begins linearly in the fields.
Writing $\phi =v+\eta $, we have%
\begin{equation*}
\langle \phi ^{\dagger }\psi \hspace{0.01in}|\hspace{0.01in}\bar{\psi}\phi
\rangle =\langle v^{\dagger }\psi \hspace{0.01in}|\hspace{0.01in}\bar{\psi}%
v\rangle +\langle \eta ^{\dagger }\psi \hspace{0.01in}|\hspace{0.01in}\bar{%
\psi}v\rangle +\langle v^{\dagger }\psi \hspace{0.01in}|\hspace{0.01in}\bar{%
\psi}\eta \rangle +\langle \eta ^{\dagger }\psi \hspace{0.01in}|\hspace{%
0.01in}\bar{\psi}\eta \rangle .
\end{equation*}%
The last three terms show that one-loop diagrams contribute to the lowest
order. In other words, the expansion in powers of the coupling does not
match the loop expansion. It might be interesting to study the 't Hooft
approach with purely virtual scalar fields $\phi $.

\bigskip

A well-known way to build manifestly gauge invariant correlation functions
of gauge fields and quarks is by means of Wilson lines. We show that, in
general, this method introduces unwanted degrees of freedom, so it cannot be
used naively to define physical absorptive parts.

We define the Wilson line $W(x,y)$ by the formula%
\begin{equation}
W(x,y)=\mathbb{P}\exp \left( ig\int_{0}^{1}\mathrm{d}s\mathcal{A}%
(x,y;s)\right) ,  \label{Wl}
\end{equation}%
where $\mathbb{P}$ denotes the path ordering and%
\begin{equation*}
\mathcal{A}(x,y;s)\equiv (x-y)^{\mu }A_{\mu }(y(1-s)+xs).
\end{equation*}%
We could consider arbitrary paths connecting $x$ to $y$, but for simplicity
we restrict to the straight segment.

We concentrate on the fermion two-point function%
\begin{equation*}
G_{\psi }(x,y)\equiv \text{Tr}\langle W(y,x)\psi (x)\bar{\psi}(y)\rangle ,
\end{equation*}%
which is clearly gauge invariant. At one loop, we can truncate the expansion
of the Wilson line to the order $g^{2}$: 
\begin{equation*}
W(x,y)=1+ig\int_{0}^{1}\mathrm{d}s\mathcal{A}(x,y;s)-g^{2}\int_{0}^{1}%
\mathrm{d}s\mathcal{A}(x,y;s)\int_{0}^{s}\mathrm{d}\tau \mathcal{A}(x,y;\tau
)+\mathcal{O}(g^{3}).
\end{equation*}%
We have 
\begin{eqnarray*}
G_{\psi }(x,y) &\equiv &\text{Tr}\langle \psi (x)\bar{\psi}(y)\rangle
+ig\int_{0}^{1}\mathrm{d}s\text{Tr}\langle \mathcal{A}(y,x;s)\psi (x)\bar{%
\psi}(y)\rangle \\
&&-g^{2}\int_{0}^{1}\mathrm{d}s\int_{0}^{s}\mathrm{d}\tau \text{Tr}\langle 
\mathcal{A}(y,x;s)\mathcal{A}(y,x;\tau )\psi (x)\bar{\psi}(y)\rangle +%
\mathcal{O}(g^{3}).
\end{eqnarray*}%
Working in momentum space, the first contribution can be read from (\ref%
{psipsi}). The second contribution is%
\begin{equation*}
\frac{i\gamma ^{\mu }p_{\mu }}{p^{2}+i\epsilon }\frac{g^{2}\lambda }{8\pi
^{2}\varepsilon }(N_{2}^{2}-1)(-p^{2}-i\epsilon )^{-\varepsilon /2},
\end{equation*}%
while the third contribution reads%
\begin{equation*}
\frac{i\gamma ^{\mu }p_{\mu }}{p^{2}+i\epsilon }\frac{g^{2}(3-\lambda )}{%
16\pi ^{2}\varepsilon }(N_{2}^{2}-1)(-p^{2}-i\epsilon )^{-\varepsilon /2}
\end{equation*}%
In total, we find%
\begin{equation*}
G_{\psi }(p)=N_{c}\frac{i\gamma ^{\mu }p_{\mu }}{p^{2}+i\epsilon }\left[ 1+%
\frac{3g^{2}}{8\pi ^{2}\varepsilon }\frac{N_{c}^{2}-1}{2N_{c}}%
(-p^{2}-i\epsilon )^{-\varepsilon /2}\right] +\mathcal{O}(g^{4}),
\end{equation*}%
the factor $N_{c}$ being due to the trace. The $\lambda $-dependence
disappears, as expected, but the absorptive part is negative, 
\begin{equation}
\text{Abso}[G_{\psi }(p)]=-\frac{3g^{2}}{2\pi }\frac{N_{c}^{2}-1}{2}%
p^{2}\theta (p^{2})+\mathcal{O}(g^{4}).  \label{Abs2}
\end{equation}%
therefore, unphysical.

\section{Conclusions}

\label{conclusions}\setcounter{equation}{0}

We have extended quantum field theory to include purely virtual
\textquotedblleft cloud\textquotedblright\ sectors, which allow us to define
gauge invariant dressed fields and study their correlation functions. The
cloud diagrammatics and its Feynman rules are derived from a local action,
built by means of cloud fields $\phi ^{i}$ and their anticommuting partners $%
H^{i}$. It includes the cloud functions, the cloud Faddeev-Popov
determinants and the cloud symmetries. The usual gauge-fixing must be cloud
invariant, while the cloud-fixings must be gauge invariant. The dressed
fields are gauge invariant, but not necessarily cloud invariant.

The extended theory is unitary, renormalizable and polynomial in all the
fields except for $\phi ^{i}$, which are dimensionless. No extra degrees of
freedom are propagated. The extension is perturbative and the expansion in
powers of the gauge coupling $g$ coincides with the expansion in the number
of loops. Each insertion in a correlation function can be equipped with its
own, independent cloud. The correlation functions of the undressed fields
are unaffected by the extra sectors. The $S$ matrix amplitudes of the
dressed fields coincide with the usual scattering amplitudes. This ensures,
among the other things, that the latter are gauge independent.

The results allow us to define short-distance scattering processes, where
the products do not have enough time to become noninteracting, asymptotic
states. Then the predictions depend on the clouds, because the observer
necessarily disturbs\ the observed phenomenon. A few initial measurements
must be sacrificed to calibrate the instrumentation. After that, everything
else is testable, and possibly falsifiable.

The dressed fields are invariant under infinitesimal gauge transformations,
but not necessarily under global gauge transformations. This is what allows
us to build physical non singlet states without violating unitarity. The
dressed fields can also be used to replace the elementary fields everywhere,
reducing the latter to mere integration and diagrammatic tools. This way,
all the calculations are manifestly gauge independent.

An extended Batalin-Vilkovisky formalism and its Zinn-Justin master
equations allow us to prove that the symmetries are preserved by
renormalization to all orders. Renormalizability by power counting is
manifest with a variety of gauge and cloud choices. With more general
choices it can be proved by means of canonical transformations.

The extra sectors propagate ghosts, and give unphysical results, if the
extra fields are quantized by means of the usual Feynman $i\epsilon $
prescription. To avoid this, those sectors are rendered purely virtual,
which means that the extra fields are quantized as fake fields. The purely
virtual nature of the cloud sectors ensures that, in the end, no unwanted
degrees of freedom propagate. This allows us to extract the physical
absorptive parts of ordinary correlation functions from the correlation
functions of the dressed fields. More generally, it opens the way to extract
physical information from off-shell correlation functions in a systematic
way.

A certain gauge/cloud duality can be used to simplify the computations. At
the conceptual level, it shows that a gauge choice is ultimately nothing but
a particular cloud, provided the gauge trivial modes are rendered purely
virtual.

We have illustrated the basic properties of the formalism by calculating the
one-loop two-point functions of the dressed quarks and gluons. Their
absorptive parts are gauge independent, cloud independent and positive.
Instead, they are unphysical if the clouds are not purely virtual (such as
those defined by the Feynman prescription). They are also unphysical,
generically speaking, if Wilson lines are used.

Among the other things, the purely virtual cloud formalism can be used as an
alternative to the popular physical gauges. Here, instead of changing the
gauge fixing to make it physical, we make a gauge-fixing physical by
changing the prescription we use for it.

The cloud fields and their partners are massless. Massless purely virtual
particles can in principle violate causality (not just microcausality) \cite%
{Absograv,classicalQG}. This aspect deserves further study. Here we just
note that the absorptive parts we have calculated are not concerned by this
fact, because they are cloud independent, so they are properties of the
fundamental theory.

\vskip 1 truecm \noindent {\large \textbf{Acknowledgments}}

\vskip .2 truecm

We are grateful to U. Aglietti, M. Bochicchio, D. Comelli, E. Gabrielli, C.
Marzo and L. Marzola for helpful discussions. This work was supported in
part by the European Regional Development Fund through the CoE program grant
TK133 and the Estonian Research Council grant PRG803. We thank the CERN
theory group for hospitality during the final stage of the project.

\vskip 1truecm

\noindent {\textbf{\huge Appendices}} \renewcommand{\thesection}{%
\Alph{section}} \renewcommand{\theequation}{\thesection.\arabic{equation}} %
\setcounter{section}{0}

\section{Notation}

\label{formulas}\setcounter{equation}{0}

In this appendix we collect the notation and some useful formulas, starting
from%
\begin{eqnarray*}
A_{\mu } &=&A_{\mu }^{a}T^{a},\qquad \lbrack
T^{a},T^{b}]=if^{abc}T^{c},\qquad \text{Tr}\left[ T^{a}T^{b}\right] =\frac{1%
}{2}\delta ^{ab},\qquad D_{\mu }\psi =\partial _{\mu }\psi -igA_{\mu }\psi ,
\\
F_{\mu \nu } &=&\partial _{\mu }A_{\nu }-\partial _{\nu }A_{\mu }-ig[A_{\mu
},A_{\nu }]=F_{\mu \nu }^{a}T^{a},\qquad F_{\mu \nu }^{a}=\partial _{\mu
}A_{\nu }^{a}-\partial _{\nu }A_{\mu }^{a}+gf^{abc}A_{\mu }^{b}A_{\nu }^{c},
\end{eqnarray*}%
where $A_{\mu }^{a}$ are the gauge fields, $F_{\mu \nu }^{a}$ is the field
strength, $T^{a}$ are the generators of the Lie group (Hermitian matrices of
the fundamental representation), $f^{abc}$ are the structure constants of
the Lie algebra, $D_{\mu }$ is the covariant derivative and $\psi $ is a
field that belongs to the fundamental representation. The gauge
transformations read%
\begin{eqnarray}
A_{\mu }^{U} &=&UA_{\mu }U^{-1}+\frac{i}{g}U\partial _{\mu }U^{-1},\qquad
F_{\mu \nu }^{U}=UF_{\mu \nu }U^{-1},\qquad \psi ^{U}=U\psi ,  \notag \\
D_{\mu }^{U} &=&UD_{\mu }U^{-1},\qquad W^{U}(x,y)=U(x)W(x,y)U^{-1}(y),
\label{gtrasf}
\end{eqnarray}%
where $U(x)=\exp \left( ig\alpha ^{a}(x)T^{a}\right) $ is a point-dependent
matrix of $SU(N_{c})$, $\alpha ^{a}(x)$ are arbitrary functions, and $W(x,y)$
is the Wilson line defined in formula (\ref{Wl}).

\section{Uniqueness of the dressed fields}

\label{uniqueness}\setcounter{equation}{0}

We prove that for every transformation law $\delta _{C}\phi ^{a}=E^{ab}(\phi
)C^{b}$ of the cloud field $\phi ^{a}$, the dressed gauge fields are unique,
up to linear combinations. Let us write%
\begin{equation}
A_{\mu \text{d}}^{a}=F^{ab}(\phi )A_{\mu }^{b}-G^{ab}(\phi )\partial _{\mu
}\phi ^{b}.  \label{adres}
\end{equation}%
If we require $(S_{K},A_{\mu \text{d}}^{a})=0$, we obtain three equations,
from the vanishing of the terms proportional to $A_{\mu }^{c}$, $\partial
_{\mu }C^{c}$ and $C^{c}\partial _{\mu }\phi ^{d}$, which are%
\begin{equation}
F^{ab}-G^{ac}E^{cb}=0,\qquad F^{ab,c}E^{cd}+gF^{ac}f^{cbd}=0,\qquad
G^{ab,c}E^{cd}+G^{ac}E^{cd,b}=0,  \label{eque}
\end{equation}%
where the comma denotes the derivative with respect to $\phi $.

We expand in powers of $g$, writing $F^{ab}=\delta ^{ab}+\sum_{n=1}^{\infty
}g^{n}F_{n}^{ab}$, $G^{ab}=\delta ^{ab}+\sum_{n=1}^{\infty }g^{n}G_{n}^{ab}$
and $E^{ab}=\delta ^{ab}+\sum_{n=1}^{\infty }g^{n}E_{n}^{ab}$. The
expansions of (\ref{eque}) to order $g^{n}$ give equations of the form%
\begin{equation*}
F_{n}^{ab}-G_{n}^{ab}=X_{n}^{ab},\qquad F_{n}^{ab,c}=Y_{n}^{abc},\qquad
G_{n}^{ab,c}=Z_{n}^{abc},
\end{equation*}%
where $X_{n}^{ab}$, $Y_{n}^{abc}$ and $Z_{n}^{abc}$ are known from the
previous orders. The ambiguity of the solution is $\Delta F_{n}^{ab}=\Delta
G_{n}^{ab}=$ constant, which can be absorbed into a redefinition $A_{\mu 
\text{d}}^{a}\rightarrow (\delta ^{ab}+g^{n}\Delta F_{n}^{ab})A_{\mu \text{d}%
}^{b}$. Thus, the most general solution $A_{\mu \text{d}}^{a}$ is a linear
combination $M^{ab}\hat{A}_{\mu \text{d}}^{b}$ of a particular solution $%
\hat{A}_{\mu \text{d}}^{b}$, where $M^{ab}$ is a constant matrix. If $%
E^{ab}(\phi )$ is an expansion is powers of $g\phi $, a unique solution has
the same property, since $M^{ab}=\delta ^{ab}$ in that case.

Similarly, the dressed fermions are unique, up to constant factors.

\section{Cloud and gauge dependence through renormalization}

\label{appe3}\setcounter{equation}{0}

In this appendix, we prove further results about the dependence of the
renormalized action on the cloud parameters, and derive some already proved
results in a more general way. See \cite{ABwardcc} for details and examples
of derivations like the ones that follow.

We add the inductive assumption that the renormalized action $S_{n\text{ tot}%
}$ to order $n$ satisfies the equation%
\begin{equation}
\frac{\partial S_{n\text{ tot}}}{\partial \tilde{\lambda}}=(S_{K}^{\text{%
cloud}},\Upsilon _{n,\tilde{\lambda}}),  \label{dr}
\end{equation}%
where $\Upsilon _{n,\tilde{\lambda}}=\partial \Upsilon _{n}/\partial \tilde{%
\lambda}$. Referring to formulas (\ref{due}), this means that $S_{n}$ does
not depend on the cloud parameters $\tilde{\lambda}$ introduced through the
cloud fermions (a result proved in a different way in subsection \ref%
{cloudindepr}). We also assume that $\langle \Upsilon _{n,\tilde{\lambda}%
}\rangle $ is convergent to the order $\hbar ^{n}$ included. Here and below
averages such as $\langle \Upsilon _{n,\tilde{\lambda}}\rangle $ are
regarded as functionals of $\Phi $, $K$, $\tilde{\Phi}$ and $\tilde{K}$.

Equation (\ref{dr}) implies%
\begin{equation}
\frac{\partial \Gamma _{n\text{ tot}}}{\partial \tilde{\lambda}}=\frac{%
\partial W_{n\text{ tot}}}{\partial \tilde{\lambda}}=\left\langle \frac{%
\partial S_{n\text{ tot}}}{\partial \tilde{\lambda}}\right\rangle =\langle
(S_{K}^{\text{cloud}},\Upsilon _{n,\tilde{\lambda}})\rangle =(S_{K}^{\text{%
cloud}},\langle \Upsilon _{n,\tilde{\lambda}}\rangle ).  \label{dgn}
\end{equation}%
The first equality follows from the definition of $\Gamma _{n\text{ tot}}$
as the Legendre transform of $W_{n\text{ tot}}$. The fourth equality can be
proved as follows (see also \cite{ABwardcc}). Consider an arbitrary action $%
S_{\text{tot}}^{\prime }$ that does not satisfy $(S_{K}^{\text{cloud}},S_{%
\text{tot}}^{\prime })=0$. Repeat the argument that leads to the proof of (%
\ref{scloudtot}) with the change of field variables 
\begin{equation*}
\Phi ^{\alpha }\rightarrow \Phi ^{\alpha },\qquad \tilde{\Phi}^{\alpha
i}\rightarrow \tilde{\Phi}^{\alpha i}+\theta (S_{K}^{\text{cloud}},\tilde{%
\Phi}^{\alpha i}),
\end{equation*}%
in the generating functional $Z_{\text{tot}}^{\prime }$ associated with $S_{%
\text{tot}}^{\prime }$. The result is 
\begin{equation}
(S_{K}^{\text{cloud}},\Gamma _{\text{tot}}^{\prime })=\langle (S_{K}^{\text{%
cloud}},S_{\text{tot}}^{\prime })\rangle ^{\prime },  \label{masterrc}
\end{equation}%
instead of (\ref{scloudtot}). The average with prime is calculated with the
action $S_{\text{tot}}^{\prime }$. Applying formula (\ref{masterrc}) to $S_{%
\text{tot}}^{\prime }=S_{n\text{\hspace{0.01in}tot}}+\theta \Upsilon _{n,%
\tilde{\lambda}}$, where $\theta $ is a constant anticommuting parameter,
and noting that $\Gamma _{\text{tot}}^{\prime }=\Gamma _{n\hspace{0.01in}%
\text{tot}}+\theta \langle \Upsilon _{n,\tilde{\lambda}}\rangle $ (which
easily follows from the definition of the $\Gamma $ functional as a Legendre
transform), we immediately get the fourth equality of (\ref{dgn}). We need
to use the identities $(S_{K}^{\text{cloud}},S_{n\text{ tot}})=(S_{K}^{\text{%
cloud}},\Gamma _{n\text{ tot}})=0$ for this. They also ensure that the
average on the right-hand side can be calculated with the unperturbed action 
$S_{n\text{ tot}}$.

Now, taking the $(n+1)$-th order divergent part of (\ref{dgn}), we obtain%
\begin{equation}
\frac{\partial \Gamma _{n\text{ tot div}}^{(n+1)}}{\partial \tilde{\lambda}}%
=(S_{K}^{\text{cloud}},\Upsilon _{n,\tilde{\lambda}\text{div}}^{(n+1)}),
\label{dr0}
\end{equation}%
where $\Upsilon _{n,\tilde{\lambda}\text{div}}^{(n+1)}$ is the $(n+1)$-th
order divergent part of $\langle \Upsilon _{n,\tilde{\lambda}}\rangle $. It
is local, since $\langle \Upsilon _{n,\tilde{\lambda}}\rangle $ is
convergent to order $n$ by assumption. Integrating (\ref{dr0}), we find%
\begin{equation*}
\Gamma _{n\text{ tot div}}^{(n+1)}=\Gamma _{n\text{ tot div}}^{(n+1)}(\tilde{%
\lambda}_{0})+\left( S_{K}^{\text{cloud}},\int_{\tilde{\lambda}_{0}}^{\tilde{%
\lambda}}\Upsilon _{n,\tilde{\lambda}\text{div}}^{(n+1)}(\tilde{\lambda}%
^{\prime })\mathrm{d}\tilde{\lambda}^{\prime }\right) ,
\end{equation*}%
where the $\tilde{\lambda}$-dependence has been emphasized in parentheses.
The second equation of (\ref{due}) then implies 
\begin{equation*}
\Gamma _{n\text{ tot div}}^{(n+1)}=\Gamma _{n\text{ div}}^{(n+1)}+(S_{K}^{%
\text{cloud}},R_{n\text{ div\hspace{0.01in}0}}^{(n+1)})+\left( S_{K}^{\text{%
cloud}},\int_{\tilde{\lambda}_{0}}^{\tilde{\lambda}}\Upsilon _{n,\tilde{%
\lambda}\text{div}}^{(n+1)}(\tilde{\lambda}^{\prime })\mathrm{d}\tilde{%
\lambda}^{\prime }\right) ,
\end{equation*}%
where $\Gamma _{n\text{ div}}^{(n+1)}$ and $R_{n\text{ div 0}}^{(n+1)}$ are $%
\tilde{\lambda}$ independent. In particular, the $\tilde{\lambda}$%
-independence of $\Gamma _{n\text{ div}}^{(n+1)}$ can be immediately proved
by taking $\tilde{\Phi}^{\alpha i}=\tilde{K}^{\alpha i}=0$. Finally, (\ref%
{subtran}) gives%
\begin{eqnarray*}
S_{n+1\text{ tot}} &=&S_{n\text{ tot}}-\Gamma _{n\text{ tot div}%
}^{(n+1)}=S_{n+1}+(S_{K}^{\text{cloud}},\Upsilon _{n+1})+S_{K}^{\text{cloud}%
}, \\
S_{n+1} &=&S_{n}-\Gamma _{n\text{ div}}^{(n+1)},\qquad \Upsilon
_{n+1}=\Upsilon _{n}-R_{n\text{ div 0}}^{(n+1)}-\int_{\tilde{\lambda}_{0}}^{%
\tilde{\lambda}}\Upsilon _{n,\tilde{\lambda}\text{div}}^{(n+1)}(\tilde{%
\lambda}^{\prime })\mathrm{d}\tilde{\lambda}^{\prime }.
\end{eqnarray*}%
We have%
\begin{equation*}
\frac{\partial S_{n+1\text{ tot}}}{\partial \tilde{\lambda}}=(S_{K}^{\text{%
cloud}},\Upsilon _{n+1,\tilde{\lambda}}),\qquad \Upsilon _{n+1,\tilde{\lambda%
}}=\Upsilon _{n,\tilde{\lambda}}-\Upsilon _{n,\tilde{\lambda}\text{div}%
}^{(n+1)}.
\end{equation*}%
Clearly, $\langle \Upsilon _{n+1,\tilde{\lambda}}\rangle $ is convergent to
the order $\hbar ^{n+1}$ included, so the inductive assumptions are
replicated to order $n+1$.\ Taking $n$ to infinity, we obtain $\Upsilon
_{R}=\Upsilon _{\infty }$ and 
\begin{equation}
\frac{\partial S_{R\hspace{0.01in}\text{tot}}}{\partial \tilde{\lambda}}%
=(S_{K}^{\text{cloud}},\Upsilon _{R,\tilde{\lambda}}).  \label{cindeptot}
\end{equation}%
Using (\ref{StotRR}) at $\tilde{\Phi}^{\alpha i}=\tilde{K}^{\alpha i}=0$, we
see that $S_{R}$ is cloud independent.

The arguments can be specialized to every cloud separately. This allows us
to prove the further identities%
\begin{equation}
\frac{\partial S_{R\hspace{0.01in}\text{tot}}}{\partial \tilde{\lambda}_{i}}%
=(S_{K}^{\text{cloud\hspace{0.01in}}i},\Upsilon _{R,\tilde{\lambda}_{i}}),
\label{cindepi}
\end{equation}%
which ensure that the $i$th cloud parameters $\tilde{\lambda}_{i}$ do not
propagate to the other cloud sectors. Moreover, formula (\ref{psiti}) tells
us that $\Upsilon _{0,\tilde{\lambda}_{i}}=\tilde{\Psi}_{i,\tilde{\lambda}%
_{i}}\equiv \partial \tilde{\Psi}_{i}/\partial \tilde{\lambda}_{i}$. The
derivation just given ensures that $\Upsilon _{R,\tilde{\lambda}_{i}}$ is
the renormalized version of the functional $\Upsilon _{0,\tilde{\lambda}%
_{i}}=\tilde{\Psi}_{i,\tilde{\lambda}_{i}}$, which we denote by $\tilde{\Psi}%
_{i,\tilde{\lambda}_{i}\hspace{0.01in}R}$.

Finally, if $X$ is a functional such that $(S_{K}^{\text{cloud\hspace{0.01in}%
}i},X)=0$, the same derivation shows that the renormalized functional $X_{R}$
satisfies $(S_{K}^{\text{cloud\hspace{0.01in}}i},X_{R})=0$. In particular, $%
(S_{K}^{\text{cloud\hspace{0.01in}}j},\Upsilon _{R,\tilde{\lambda}_{i}})=0$
for every $j\neq i$.

Similar procedures (see \cite{ABwardcc}) allow us to prove 
\begin{equation}
\frac{\partial S_{R\hspace{0.01in}\text{tot}}}{\partial \tilde{\lambda}_{i}}%
=(S_{R\hspace{0.01in}\text{tot}},\tilde{\Psi}_{i,\tilde{\lambda}_{i}\hspace{%
0.01in}R}),\qquad \frac{\partial S_{R\hspace{0.01in}\text{tot}}}{\partial
\lambda }=(S_{R\hspace{0.01in}\text{tot}},\Psi _{\lambda \hspace{0.01in}%
R}),\qquad (S_{K}^{\text{cloud\hspace{0.01in}}i},\Psi _{\lambda \hspace{%
0.01in}R})=0,  \label{cindepR}
\end{equation}%
where $\Psi _{\lambda \hspace{0.01in}R}$ is the renormalized version of the
functional $\Psi _{\lambda \hspace{0.01in}}=\partial \Psi /\partial \lambda $%
.

Summarizing the cloud dependence is encoded in the equations%
\begin{equation}
\frac{\partial S_{R\hspace{0.01in}\text{tot}}}{\partial \tilde{\lambda}_{i}}%
=(S_{R\hspace{0.01in}\text{tot}},\tilde{\Psi}_{i,\tilde{\lambda}_{i}\hspace{%
0.01in}R}),\qquad \frac{\partial S_{R\hspace{0.01in}\text{tot}}}{\partial 
\tilde{\lambda}_{i}}=(S_{K}^{\text{cloud\hspace{0.01in}}i},\tilde{\Psi}_{i,%
\tilde{\lambda}_{i}\hspace{0.01in}R}),\qquad (S_{K}^{\text{cloud\hspace{%
0.01in}}j},\tilde{\Psi}_{i,\tilde{\lambda}_{i}\hspace{0.01in}R})=0\text{ for 
}j\neq i,  \label{cind}
\end{equation}%
while the gauge dependence is encoded in%
\begin{equation}
\frac{\partial S_{R\hspace{0.01in}\text{tot}}}{\partial \lambda }=(S_{R%
\hspace{0.01in}\text{tot}},\Psi _{\lambda \hspace{0.01in}R}),\qquad (S_{K}^{%
\text{cloud\hspace{0.01in}}i},\Psi _{\lambda \hspace{0.01in}R})=0.
\label{gind}
\end{equation}%
Integrating the first equations of (\ref{cind}) and (\ref{gind}) as shown in 
\cite{ABwardcc}, it is straightforward to prove that the dependence on the
cloud parameters is a canonical transformation and the dependence on the
gauge fixing parameters is also a canonical transformation.

\end{document}